\pdfoutput=1
\documentclass[sigconf]{acmart}

\usepackage{adjustbox}
\usepackage{appendix}
\usepackage{array}
\usepackage{booktabs}
\usepackage{colortbl}
\usepackage{csquotes}
\usepackage{cuted}
\usepackage[shortlabels]{enumitem}
\usepackage{graphicx}
\usepackage{multirow}
\usepackage{subcaption}
\usepackage{tabularx}
\usepackage{tcolorbox}
\usepackage{ulem}
\usepackage{xcolor}
\usepackage{xtab}
\usepackage{xurl}

\tcbset{before upper={\parindent0em}}
\newtcolorbox[auto counter]{quotebox}{boxsep=-5pt,colback=gray!20,width=1.05\columnwidth}

\newcounter{myquote}
\renewcommand{\themyquote}{\ref{sec:appendix:quotes}\arabic{myquote}}
\makeatletter
\newenvironment{myquote}[1]{
\refstepcounter{myquote}
\noindent
\vspace{-0.25em}
\begin{adjustbox}{width=0.995\columnwidth}
\begin{quotebox}
\hspace{-0.05\columnwidth} 
\begin{minipage}[l]{\columnwidth}
\noindent
\begin{displayquote}[\textit{#1}]}
{\end{displayquote} \end{minipage}%
\hspace{0.025\columnwidth} 
\begin{minipage}[r]{0.1\columnwidth} \noindent (\themyquote) \end{minipage}%
\end{quotebox}
\end{adjustbox}
\vspace{-0.25em}}

\newcolumntype{s}{>{\columncolor[rgb]{0.9,0.9,0.9}}c}

\AtBeginDocument{%
  \providecommand\BibTeX{{%
    \normalfont B\kern-0.5em{\scshape i\kern-0.25em b}\kern-0.8em\TeX}}}

\copyrightyear{2021}
\acmYear{2021}
\setcopyright{rightsretained}
\acmConference[WPES '21] {Proceedings of the 20th Workshop on Privacy in the Electronic Society}{November 15, 2021}{Virtual Event, Republic of Korea.}
\acmBooktitle{Proceedings of the 20th Workshop on Privacy in the Electronic Society (WPES '21), November 15, 2021, Virtual Event, Republic of Korea}
\acmPrice{}
\acmISBN{978-1-4503-8527-5/21/11}
\acmDOI{10.1145/3463676.3485601}
\begin{document}
\fancyhead{}
\title{Fighting the Fog: Evaluating the Clarity of Privacy Disclosures in the Age of CCPA}

\author[1]{Rex Chen}
\affiliation{\institution{Institute of Software Research, \\School of Computer Science, \\Carnegie Mellon University}}
\email{rexc@cmu.edu}

\author[2]{Fei Fang}
\affiliation{\institution{Institute of Software Research, \\School of Computer Science, \\Carnegie Mellon University}}
\email{feifang@cmu.edu}

\author[3]{Thomas Norton}
\affiliation{\institution{Center on Law and Information Policy (CLIP), School of Law, \\Fordham University}}
\email{tnorton1@fordham.edu}

\author[4]{Aleecia M. McDonald}
\affiliation{\institution{Institute of Software Research, \\School of Computer Science, \\Carnegie Mellon University}}
\email{am40@andrew.cmu.edu}

\author[5]{Norman Sadeh}
\affiliation{\institution{Institute of Software Research, \\School of Computer Science, \\Carnegie Mellon University}}
\email{sadeh@cs.cmu.edu}

\renewcommand{\shortauthors}{Chen et al.}

\begin{abstract}
Vagueness and ambiguity in privacy policies threaten the ability of consumers to make informed choices about how businesses collect, use, and share their personal information. The California Consumer Privacy Act (CCPA) of 2018 was intended to provide Californian consumers with more control by mandating that businesses (1) clearly disclose their data practices and (2) provide choices for consumers to opt out of specific data practices. In this work, we explore to what extent CCPA's disclosure requirements, as implemented in actual privacy policies, can help consumers to answer questions about the data practices of businesses. First, we analyzed 95 privacy policies from popular websites; our findings showed that there is considerable variance in how businesses interpret CCPA's definitions. Then, our user survey of 364 Californian consumers showed that this variance affects the ability of users to understand the data practices of businesses. Our results suggest that CCPA's mandates for privacy disclosures, as currently implemented, have not yet yielded the level of clarity they were designed to deliver, due to both vagueness and ambiguity in CCPA itself as well as potential non-compliance by businesses in their privacy policies.
\end{abstract}




\maketitle

\section{Introduction}
\label{sec:intro}
Privacy policies are generally difficult to understand. Most privacy policies contain vague or ambiguously-worded statements, for which even legal experts often cannot agree on the correct interpretations \citep{Reidenberg2015}. For consumers, vague and ambiguous privacy policies pose a risk: a fundamental tenet of consumer privacy regulation in the United States is that consumers must be able to make informed decisions about the collection and use of their data, and clear privacy policies are the primary mechanism that provides consumers with the information necessary to make these decisions \cite{Schwartz2009}.

One of the main goals of the California Consumer Privacy Act (CCPA) of 2018 has been to introduce more stringent requirements that provide consumers with ``more control over the personal information that businesses collect about them'' \citep{OAG,Field2020}. CCPA provides consumers with the rights (1) to request that a business provide specific disclosures regarding the use and sharing of personal information collected from them; (2) to request that a business not sell their personal information by opting out of sales; (3) to request that a business delete personal information that it has collected about them; and (4) to not be discriminated against based on exercising their CCPA privacy rights \citep{Kessler2019}. One year after CCPA went into force, we explore how businesses have responded to its requirements by providing the first systematic assessment focusing on the clarity of their CCPA-mandated privacy disclosures.

To this end, we analyzed the privacy policies of 95 of the most popular websites in the United States. We found that the legal terms introduced by CCPA have been used inconsistently by different businesses. For example, while many businesses state that they do not ``sell'' personal information, others with ostensibly similar practices acknowledge that their sharing of information with third parties may fall under CCPA's definition of a ``sale''. Different CCPA disclosures include also inconsistent levels of detail about what categories of personal information are shared with what categories of recipients; what categories of personal information may be sold; and what outcomes consumers may expect for requests to delete their personal information. If businesses do not consistently interpret CCPA's requirements, consumers also cannot be reasonably expected to consistently interpret these businesses' privacy policies. 

We also conducted a user survey ($n = 364$) to evaluate how the wording of disclosures affects consumers' understanding of three potential ambiguities that we identified in our privacy policy analysis: (1) the recipients of specific categories of shared personal information are often specified vaguely; (2) the extent to which opt-out requests can limit the sharing of personal information with third parties is often unclear; and (3) the purposes of retaining personal information after a successful deletion request, as well as potential disclosures of this information, are often undefined. We showed respondents excerpts from privacy policies that reflect a substantial range of existing patterns in disclosure content. Then, we asked them to evaluate the likelihood of data usage scenarios given the excerpts, and we compared their responses with our ground truth answers. Our results show that, separately from policy-level usability (\textit{cf.} \cite{Vu2007}), respondents' understandings of disclosures suffered from inconsistent usage of terminology along with a lack of knowledge about how terms such as ``sale'', ``business purpose'', and ``service provider'' are to be interpreted under CCPA.

Our results suggest that CCPA requirements for privacy disclosures, as currently implemented, have not sufficiently eliminated vagueness and ambiguity in privacy policies. This is in part due to a lack of specificity in some requirements, differences in interpretations of the requirements, as well as limitations on the enforcement actions taken so far. Our results suggest a need for further clarification and refinement of disclosure requirements to eliminate sources of vagueness and misinterpretation, through revisions such as the recently-passed California Privacy Rights Act (CPRA), in combination with more systematic enforcement activities.

\section{Background}
\label{sec:bg}
\subsection{The ``notice-and-choice'' paradigm}
\label{sec:bg:notice}
In the United States, the regulation of consumer privacy has historically been based on the \emph{notice-and-choice} paradigm: businesses are responsible for providing consumers with \emph{notices} about how their data is collected, shared, and used, as well as \emph{choices} for these practices \cite{Reidenberg2015}. This paradigm was founded on the Fair Information Practice Principles (FIPPs) described in a 1973 report by the Department of Health, Education and Welfare and adopted by the Federal Trade Commission in 1998 \cite{Cate2006}, as well as the privacy guidelines established by the Organization for Economic Co-operation and Development (OECD) \cite{McDonald2008}. The United States has taken a market-driven approach to privacy regulation: businesses voluntarily provide notice using \emph{privacy policies}, but there are few regulations about the format, length, and readability of privacy policies \cite{Schwartz2009}.

Notice-and-choice has been considered ineffective \cite{Cate2010,Feng2021,Schaub2015}. A primary concern has been the lack of transparency in privacy policies \cite{Reidenberg2015,Schwartz2009}. Businesses have no incentive to make specific disclosures, lest they restrict potential future usage of their data \cite{Schwartz2009}. Legally speaking, the non-disclosure of a practice does not mean that a business is prohibited from performing it \cite{Reidenberg2015}, which incentivizes businesses to use ambiguous language. Another major issue has been the length of privacy policies. \cite{McDonald2008} estimated that a typical Internet user who carefully reads the privacy policies on the websites that they frequent would need to spend, on average, 244 hours doing so annually. Furthermore, consumers usually have few choices for the entities that they can do business with; combined with the ``take it or leave it'' approach to privacy controls \cite{Contissa2018} that has been adopted by many businesses, this implies that the element of choice in notice-and-choice is ultimately illusory \cite{Kaminski2020,Schwartz2009}.

\subsection{The California Consumer Privacy Act}
\label{sec:bg:ccpa}
Owing to a 1972 ballot measure, California is now among the ten US states where privacy is enshrined as a constitutional right \cite{Kelso1992}. In 2004, the California Online Privacy Protection Act (CalOPPA), then the most expansive privacy law in the US, came into effect. Yet, it was still focused on the ``notice'' aspect of the notice-and-choice paradigm \cite{Pardau2018}. On June 28, 2018, Californian legislators passed the California Consumer Privacy Act (CCPA), which provided consumers with rights for both notice and choice (see Section \ref{sec:bg:ccpa:content}). Other states have since passed laws that are similar to CCPA \cite{Kaminski2020}.

CCPA has been considered to be inspired by the European Union General Data Protection Regulation (GDPR) (\textit{e.g.}, \cite{Field2020}). In particular, the opt-out and deletion choices in CCPA and its broad applicability across industry sectors are more reminiscent of GDPR than prior sectoral US legislation. However, fundamental differences show that CCPA and GDPR are still grounded in divergent approaches toward privacy \cite{Chander2019,Park2019}. Specifically, CCPA places an emphasis on \emph{opting out}, \textit{i.e.} companies may collect personal information about adult consumers until they withdraw their consent. Meanwhile, GDPR places an emphasis on \emph{opting in}, \textit{i.e.} companies may not collect or process personal information without either obtaining consent from the consumer or specifying a valid legal basis \cite{Chander2019,Baik2020}. (For children younger than 16, CCPA also requires opt-in consent.)

\subsubsection{Provisions}
\label{sec:bg:ccpa:content}
CCPA includes the following provisions for the privacy rights of consumers who are adult California residents, which provide notice and control about the collection and use of their personal information by businesses operating in California:
\begin{itemize}
    \item \textbf{The right to know}: Consumers have the right to request that businesses disclose the ``categories of personal information'' they have collected; ``the categories of sources'' from which the information was collected; ``the business or commercial purpose for collecting or selling'' information; ``the categories of third parties'' with whom information has been shared; and ``the specific pieces of information'' that have been collected (California Civil Code \S 1798.110(a)). In response to a verifiable request from a consumer, a business must provide this information free of charge based on its activity in the past 12 months (\S 1798.130(a)(2)). Businesses must also disclose, ``at or before the point of collection'', the categories of personal information that they collect and the purposes for which it will be used (\S 1798.100(a)).
    
    \item \textbf{The right to delete}: Consumers have the right to request that businesses delete \emph{any} personal information that has been collected from them (\S 1798.105(a)). Businesses will be required to comply with deletion requests except when the information is used for one of several \textbf{business purposes}, including legal, security, and research activities, but also other commercial and internal activities that are ``reasonably anticipated within the context of a business's ongoing business relationship with the consumer'' (\S 1798.105(d)).
    
    \item \textbf{The right to opt out of sale}: Consumers have the right to direct businesses that sell their personal information to not sell their personal information (\S 1798.120(a)). Businesses must provide notice that they sell personal information and that consumers have the right to opt out (\S 1798.120(b)), and they must respect opt-out requests for at least 12 months before asking the consumer again (\S 1798.135(a)(5)). CCPA defines the term ``sell'' as ``selling, renting, releasing, disclosing, disseminating, making available, transferring, or otherwise communicating [...] a consumer's personal information by the business to another business or a third party for monetary or other valuable consideration'' (\S 1798.140 (t)(1)). For consumers under 16 years of age, businesses must obtain opt-in consent (\S 1798.120(c)) before selling information.
    % Businesses that sell personal information must also add a conspicuous link on their website home page entitled ``Do Not Sell My Personal Information'', which allows users to opt out (\S 1798.135(a)). 
    
    \item \textbf{The right to non-discrimination}: Businesses are not permitted to discriminate against consumers who exercise the previously-defined rights, including by denying goods or services; charging different prices; providing different levels of quality; or suggesting that the price or level of quality provided will be different (\S 1798.125(a)), except if the differences are ``reasonably related to the value provided to the consumer by the consumer's data'' (\S 1798.125(b)).
\end{itemize}

We now expand on several terms and definitions in CCPA that are relevant to the scope of our study.
\begin{itemize}
    \item \textbf{Personal information} is defined as information that identifies or can be reasonably linked with individual consumers or households, including one of twelve categories: identifiers (\textit{e.g.}, name, postal address, email address, etc.); protected classifications under state or federal law (\textit{e.g.}, gender, race, religion, etc.); commercial information (\textit{e.g.}, purchase records, histories, or tendencies); biometric information; internet or other electronic network activity information (\textit{e.g.}, browsing or search history, or activity on a website); geolocation; sensory information (\textit{e.g.}, audio, electronic, visual, thermal, olfactory, etc.); professional or employment-related information; non-publicly available education information; and inferences drawn from the preceding categories of personal information (\S 1798.140(o)). Responses to right-to-know requests must follow this categorization (\S 1798.130(a)).
    
    \item A \textbf{service provider} is a for-profit legal entity ``that processes information on behalf of a business and to which the business discloses a consumer's personal information for a business purpose pursuant to a written contract'', provided that the contract binds the service provider from using the information for purposes outside of the contract (\S 1798.140(v)). Businesses must direct their service providers to delete the personal information of consumers in response to deletion requests that they have received (\S 1798.105(c)).
    
    \item \textbf{Verifiable consumer requests} refer to requests for which the business can reasonably verify that the requester is the consumer ``about whom the business has collected personal information'' (or an authorized representative, or a parent if the consumer is a minor). Businesses are not obliged to respond to requests if they are unable to verify the identity of the consumer or their representative (\S 1798.140(y)).
\end{itemize}

\subsubsection{Expansions of CCPA}
\label{sec:bg:ccpa:expansion}
After CCPA came into effect on January 1, 2020, it was augmented with a set of regulations approved on August 14, 2020, which provide further guidance to businesses on how to implement CCPA's mandates \cite{OAGAug2020}. Several modifications to the regulations have been proposed, with the fourth set of proposed modifications having been most recently released on December 14, 2020. These modifications went into effect on March 15, 2021 \cite{OAGMar2021}. We now highlight some of the currently implemented regulations that are relevant to the scope of our study.

\begin{itemize}
    \item With respect to notices made by businesses at the point of collection of personal information, the descriptions of the categories of information collected must provide consumers with ``a meaningful understanding of the information being collected'' (California Code of Regulations \S 999.305(b)). ``For each category of personal information identified'', the privacy policy text must list ``the categories of third parties to whom the information was disclosed or sold'' for business purposes. (CCPA and CCPA regulations currently use ``share'' and ``disclose'' interchangeably; we adopt the former terminology to avoid confusion.) Privacy policies must also give notice of the right to know. (\S 999.308(c)(1))
    
    \item The privacy policy text must explain whether the business sells personal information, and that the user has the right to opt-out of the sale of their personal information regardless of whether the business sells or does not sell personal information (\S 999.308(c)(3)). However, if the business does not sell personal information, a notice of this right separate from the privacy policy is not needed (\S 999.306(d)).
    
    \item The privacy policy text must also explain consumers' right to deletion, along with instructions to submit verifiable deletion requests and a description of the process by which the business verifies deletion requests (\S 999.308(c)(2)).
    
    % \item Service providers must not sell data ``on behalf of a business'' when consumers have opted out of the sale of their information with the business (\S 999.314(d)).
\end{itemize}

Concurrently, the California Privacy Rights Act (CPRA) \cite{CPRA2020} introduced an expansion of CCPA. CPRA was proposed as a new ballot initiative and approved as California Proposition 24 on November 3, 2020. However, it will not go into effect until January 1, 2023. CPRA will delegate enforcement of consumer privacy to a new, independent agency, the California Privacy Protection Agency, instead of the California Attorney General \cite{NAI2020}. In Section \ref{sec:conclusion:policy}, we will consider the changes introduced by CPRA in light of our results concerning vagueness and ambiguity in privacy policies.

\subsubsection{Criticism of CCPA}
\label{sec:bg:ccpa:criticism}
In an effort to replace a stricter ballot initiative that would have been more difficult to modify, the passage of CCPA was rushed through both houses of the California State Legislature \cite{Pardau2018}. This resulted in ``aggressive language and ambiguous terms'' in the version of the Act that was initially passed \cite{Li2019}. The concerns that have been raised over CCPA are twofold: businesses have raised concerns about CCPA's practical implementability, while consumer advocates have raised concerns about the clarity and strength of CCPA's provisions. Both groups have drawn attention to errors in CCPA that require updates. \cite{Baik2020,Li2019}

Corporate representatives have highlighted the implementation costs of CCPA compliance, especially to smaller businesses (although some exemptions apply) \cite{Li2019}. This issue is exacerbated by ambiguities in CCPA's definition of the terms ``sale'' (especially the term ``valuable consideration'' in the definition), ``consumer'', ``personal information'', ``inferences'', ``publicly available information'', and ``business purpose'' \cite{Illman2020,Li2019,Viljoen2020}, which has led to confusion on the scope of necessary compliance actions. Such ambiguities do not just impact businesses: consumers are placed at risk when businesses adopt more restrictive interpretations of these terms than they do \cite{Spivak2020}. A comparison of corporate and consumer public comments on CCPA showed that this was the case \cite{Baik2020}. 

The exemptions provisioned by CCPA have also been viewed as problematic. For instance, the right to delete may be undermined by the exemption for ``reasonably anticipated'' internal activities, and the non-discrimination clause may allow businesses to determine the ``value'' of personal information in a way that effectively permits discrimination against consumers \cite{Spivak2020,Viljoen2020}.

\section{Privacy policy analysis}
\label{sec:policies}
To understand how businesses have interpreted and implemented CCPA's requirements in light of the legal ambiguities mentioned in Section \ref{sec:bg:ccpa:criticism}, we analyzed the language and formatting used in the privacy policies of a sample of popular websites. We believe that such policies will be representative of those typically encountered by consumers because the popularity of websites follows a long-tailed power law distribution \cite{Krashakov2006,Pochat2019}. Specifically, we analyzed the privacy policies of 95 out of the 100 most popular websites in the United States (excluding 5 adult websites, as adult websites have access patterns and demographics distinct from other websites \cite{Ahmed2016}), as ranked by Alexa \cite{Alexa2021} on February 15, 2021. Appendix \ref{sec:appendix:archive} includes links to Internet Archive snapshots of the policies as they appeared on the date of the analysis. For policies that contained sections addressing the disclosure requirements of CCPA regulations (\S 999.308(c)), we focused our analysis on these sections, along with any other sections that were referenced by this text. If the policy did not contain such sections, we analyzed the entire policy.

% In the following, we summarize the broad patterns that we observed.

We performed de-duplication on our sample of 95 privacy policies by identifying all sets of websites that linked to the same privacy policies in their footers. This left 86 unique privacy policies, 67 of which include one or more sections dedicated to CCPA disclosures. The 19 unique privacy policies without CCPA-specific disclosures include those for government organizations (the USPS, the California government, the IRS, and the NIH); foreign businesses (Tmall, TikTok, QQ, Sohu, and Alibaba, based in China; the BBC, based in the UK); and entities that collect minimal personal information (Wikipedia, DuckDuckGo). Other privacy policies without CCPA-specific disclosures do not fall under these exemptions, but have mostly incorporated the required disclosures into the text (Shopify, Dropbox, Twitter, Heavy, TradingView, WordPress, and CNBC).

Overall, we observed substantial variation in the content and structure of CCPA-mandated disclosures. Except for passages of text that are based on or reference the language of CCPA, these disclosures are largely tailored to the specific activities and data practices of each business. This is a positive signal that these businesses have not relied on boilerplate text. Yet, the variability we observed also extends to businesses' interpretations of CCPA's terms and regulations. Descriptions of consumers' privacy rights or businesses' data practices in these privacy policies are often vague and ambiguous, which may reflect CCPA's underlying lack of clarity.

% We also briefly note that the CCPA-mandated disclosures for these websites are often difficult to access or comprehend because they are spread across multiple sections. In 51 unique privacy policies, CCPA-specific disclosures are not self-contained: users need to navigate to other sections of the privacy policy to find full details regarding at least one of the three types of privacy disclosures that we consider. Yet, as in the formatted Excerpt \ref{quote:zoom}, references to other sections in CCPA-specific disclosures are also not always hyperlinked. We do not focus on this issue in the rest of our work.

After an initial exploratory analysis, we selected three types of privacy disclosures that permit comparative analysis in the context of past work on GDPR: (1) the right to know, along with the categories and recipients of personal information that the business has shared (California Code of Regulations \S 999.308(c)(1)); (2) the right to opt out of sale and whether the business sells personal information (\S 999.308 (c)(3)); and (3) the right to deletion and the deletion request process (\S 999.308 (c)(2)). Based on our exploratory analysis, we also developed a list of patterns of interest for each disclosure type, which we applied to the privacy policy excerpts in two subsequent rounds of coding. Privacy policy excerpts that exemplify the patterns we observed are listed in Appendix \ref{sec:appendix:quotes}.

\subsection{Sharing of personal information}
\label{sec:policies:sharing}
\begin{table*}[ht]
\aboverulesep=0ex
\belowrulesep=0ex
    \centering
    \begin{tabular}{@{}c|cssss|c@{}}
        \toprule
        \rowcolor[rgb]{1,1,1}
        \multirow{2}{*}{\textbf{Purpose association}} & \multicolumn{5}{c|}{\textbf{Personal information/recipient category association}} & \multirow{2}{*}{\textbf{Total}} \\
        & \textit{Mapping} & \textit{Examples by PI} & \textit{Examples by recipient} & \textit{Separate lists} & \textit{Shared PI missing} & \\
        \midrule
        \textit{Mapping by PI} & 10 (12.35\%) & 0 (0\%) & 0 (0\%) & 1 (1.23\%) & 2 (2.47\%) & 13 (16.05\%) \\
        \textit{Examples by PI} & 1 (1.23\%) & 1 (1.23\%) & 3 (3.7\%) & 0 (0\%) & 0 (0\%) & 5 (6.17\%) \\
        \textit{Mapping by recipient} & 3 (3.7\%) & 0 (0\%) & 9 (11.11\%) & 9 (11.11\%) & 6 (7.41\%) & 27 (33.33\%) \\
        \textit{Examples by recipient} & 2 (2.47\%) & 3 (3.7\%) & 4 (4.94\%) & 6 (7.41\%) & 6 (7.41\%) & 21 (25.93\%) \\
        \textit{Separate lists} & 4 (4.94\%) & 0 (0\%) & 1 (1.23\%) & 8 (9.88\%) & 3 (3.7\%) & 16 (19.75\%) \\
        \midrule
        \textbf{Total} & 20 (24.69\%) & 4 (4.94\%) & 17 (20.99\%) & 24 (29.63\%) & 17 (20.99\%) & 82 \\
        \bottomrule
    \end{tabular}
    \medskip
    \caption{Summary of patterns in sharing disclosures across all 82 unique privacy policies with such disclosures. See text of Section \ref{sec:policies:sharing} for pattern definitions. Percentages show the proportion of all policies with sharing disclosures that exhibit a given pattern or combination of patterns. Shaded cells denote possible non-compliance. PI stands for ``personal information''.}
    \label{tab:sharing}
\vspace{-2em}
\end{table*}

All but four businesses (which lack CCPA-specific disclosures or do not collect personal information) describe their sharing practices in their privacy policies. Among the descriptions of sharing practices in 82 unique privacy policies, we found inconsistent levels of specificity regarding the categories of personal information that businesses share and the categories of recipients that they share this information with. CCPA regulations stipulate that the recipients for \emph{each} category of shared personal information must be specified individually (\S 999.308(c)(1)). We observed the following patterns with respect to this association, which are summarized in Table \ref{tab:sharing}.

\begin{itemize}
    \item \textbf{One-to-one mapping} (Excerpt \ref{quote:microsoft}): Following CCPA regulations, the text exhaustively lists (\textit{i.e.}, does not indicate that any list consists of examples) every category of personal information that is shared with each category of recipients.
    
    \item \textbf{Examples by PI category} (Excerpt \ref{quote:amazon}): The text exhaustively lists categories of personal information, but indicates that the lists of recipients associated with these categories are incomplete (by ``for example'', ``such as'', or similar).
    
    \item \textbf{Examples by recipient category} (Excerpt \ref{quote:nyt}): The text exhaustively lists categories of recipients or sharing purposes, but indicates that the associated lists of personal information are incomplete (by ``for example'', ``such as'', or similar).
    
    \item \textbf{Separate lists} (Excerpt \ref{quote:instructure}): The text does not associate categories of personal information and categories of recipients, or it uninformatively specifies that each category of personal information is shared with \emph{all} recipients (or vice versa).
    
    \item \textbf{No list of shared personal information categories} (Excerpt \ref{quote:fedex}): The text does not include a list of personal information that is \emph{shared}, but instead generically refers to ``personal information'' in its description of sharing practices.
\end{itemize}

Based on CCPA regulations, we would expect all of the sharing disclosures to fall into the first category; yet, 75.61\% of disclosures do not. Particularly troublesome are those that make no category-based association whatsoever, including those that only vaguely refer to ``personal information''. The latter disclosures invariably list the categories of personal information that are \emph{collected} by these businesses --- a reasonable inference is that all categories of the personal information they collect may be shared, but this cannot be confirmed from the text of these disclosures.

This lack of clarity is illustrative of the vagueness often found in privacy policies. If privacy policies do not clearly associate categories of shared personal information with their recipients, consumers cannot determine the extents to which these different categories of information are shared with different recipients. Such vagueness can be equated to a form of ``dark pattern'' \cite{Bosch2016}, as it can lead consumers to misconstrue the text of disclosures such that they fail to recognize the possibility of undesirable sharing practices. Our user survey results in Section \ref{sec:survey-results} echo this theme.

We also examined disclosures regarding the business purposes for which information is shared, which are not mandatory under CCPA. The following patterns are also summarized in Table \ref{tab:sharing}:

\begin{itemize}
    \item \textbf{Mapping by PI category} (Excerpt \ref{quote:microsoft}): The text exhaustively defines purposes of sharing personal information, associating them with categories of personal information.
    
    \item \textbf{Examples by PI category} (Excerpt \ref{quote:amazon}): The same, but the text indicates that the list of purposes is incomplete.
    
    \item \textbf{Mapping by recipient category} (Excerpt \ref{quote:salesforce}): The text exhaustively defines purposes of sharing personal information, associating them with categories of recipients.
    
    \item \textbf{Examples by recipient category} (Excerpt \ref{quote:reddit}): The same, but the text indicates that the list of purposes is incomplete.
    
    \item \textbf{Generic list} (Excerpt \ref{quote:fedex}): The text does not associate purposes of sharing with personal information or recipients.
\end{itemize}

Interestingly, across the different types of associations between personal information and recipient categories, we found that the specificity of disclosures about the purposes of sharing personal information did not differ appreciably. Even 65.85\% of policies that provide no such association still organize their purposes of sharing by categories of recipients. We speculate that these disclosures are more complete because GDPR has stringent mandates regarding the ``legal bases'' of processing personal information \cite{Gonzalez2019,Contissa2018}.

Based on prior analyses of privacy policies within the framework of GDPR, the patterns of vagueness we observed are not novel: the automated analysis of \cite{Linden2020} identified a large proportion of privacy policies that are insufficiently specific about the categories of shared personal information, and \cite{Mohan2019} specifically pointed to several privacy policies with unclear disclosures about sharing practices. Yet, our result is still surprising given that CCPA has \emph{more stringent} requirements: GDPR does not require categories of personal information to be mapped to recipients \cite{Contissa2018}. The broad pattern of apparent non-compliance in 75.61\% of disclosures suggests that inadequate enforcement may contribute to the status quo.

\subsection{Sale of personal information}
\label{sec:policies:sale}
\begin{table*}[ht]
\aboverulesep=0ex
\belowrulesep=0ex
    \centering
    \begin{tabular}{@{}c|c|ccccs|c@{}}
        \toprule
        \rowcolor[rgb]{1,1,1}
        & & \multicolumn{5}{c|}{\textbf{Interpretation of ``sale''}} & \multirow{2}{*}{\textbf{Total}} \\
        & & \textit{No sale} & \textit{Advertising} & \textit{Sharing (other)} & \textit{No reframing} & \textit{Silence} & \\
        \midrule
        \multirow{3}{*}{\textbf{Justification}} & \textit{Operational} & 13 (15.12\%) & 12 (13.95\%) & 9 (10.47\%) & 2 (2.33\%) & N/A & 36 (41.86\%) \\
        & \textit{Legal} & 13 (15.12\%) & 2 (2.33\%) & 3 (3.49\%) & 0 (0\%) & N/A & 18 (20.93\%) \\
        & \textit{None} & 21 (24.42\%) & N/A & N/A & 1 (1.16\%) & 10 (11.63\%) & 32 (37.21\%) \\
        \midrule
        \multirow{2}{*}{\textbf{Disclosure of advertising}} & \textit{Direct} & 26 (30.23\%) & 14 (16.28\%) & 10 (11.63\%) & 2 (2.33\%) & 6 (6.98\%) & 58 (67.44\%) \\
        & \textit{Indirect} & 8 (9.3\%) & 0 (0\%) & 2 (2.33\%) & 1 (1.16\%) & 2 (2.33\%) & 13 (15.12\%) \\
        \midrule
        \multirow{4}{*}{\textbf{Opt out of sales}} & \textit{Provided} & 3 (3.49\%) & 14 (16.28\%) & 12 (13.95\%) & 3 (3.49\%) & 1 (1.16\%) & 33 (38.37\%) \\
        & \textit{Hypothetical} & 2 (2.33\%) & N/A & N/A & N/A & 2 (2.33\%) & 4 (4.65\%) \\
        & \textit{Denied} & 6 (6.98\%) & N/A & N/A & N/A & 0 (0\%) & 6 (6.98\%) \\
        \rowcolor[rgb]{0.9,0.9,0.9}
        \cellcolor[rgb]{1,1,1} & \textit{Silence} & 36 (41.86\%) & 0 (0\%) & 0 (0\%) & 0 (0\%) & 7 (8.14\%) & 43 (50\%) \\
        \midrule
        \multirow{3}{*}{\textbf{Categorization of PI}} & \textit{Full} & N/A & 8 (9.3\%) & 9 (10.47\%) & 1 (1.16\%) & N/A & 18 (20.93\%) \\
        & \textit{Partial} & N/A & 4 (4.65\%) & 2 (2.33\%) & 0 (0\%) & N/A & 6 (6.98\%) \\
        & \textit{None} & 47 (54.65\%) & 2 (2.33\%) & 1 (1.16\%) & 2 (2.33\%) & 10 (11.63\%) & 62 (72.09\%) \\
        \midrule
        \multicolumn{2}{c|}{\textbf{Total}} & 47 (54.65\%) & 14 (16.28\%) & 12 (13.95\%) & 3 (3.49\%) & 10 (11.63\%) & 86 \\
        \bottomrule
    \end{tabular}
    \medskip
    \caption{Summary of patterns in sale disclosures across all 86 unique privacy policies. See text of Section \ref{sec:policies:sale} for pattern definitions. Percentages show the proportion of all policies that exhibit a given pattern or combination of patterns. Shaded cells denote possible non-compliance. PI stands for ``personal information''.}
    \label{tab:sale}
\vspace{-2em}
\end{table*}

We found that disclosures in privacy policies about the right to opt out of sale are fundamentally inconsistent in their varying interpretations of CCPA's definition of a ``sale''. We identified five types of interpretations, which are summarized in Table \ref{tab:sale}:

\begin{itemize}
    \item \textbf{No sale} (Excerpt \ref{quote:spotify}, \ref{quote:linkedin}): The text states directly that the business does not sell personal information.
    
    \item \textbf{Reframed as sharing} (Excerpt \ref{quote:cnn}): The text states that some of the business' sharing practices may fall under the definition of a ``sale'', but not in the traditional sense.
    
    \begin{itemize}
        \item \textbf{Reframed as advertising} (Excerpt \ref{quote:hulu}): The text primarily (\textit{i.e.}, as the longest example or the first example) refers to the business' advertising or marketing activities as being possibly interpretable as ``sales''.
    \end{itemize}
    
    \item \textbf{No reframing} (Excerpt \ref{quote:target}, \ref{quote:healthline}): The text states that the business sells personal information without qualifying such statements by appealing to an alternate definition.
    
    \item \textbf{Silence}: The text does not state whether the business sells any kind of personal information.
\end{itemize}

While these patterns could reflect genuine variation in practice, our analysis suggested that this may not be the case: the justifications provided by businesses for their interpretations of CCPA's definition are also inconsistent. This suggests that the latitude provided by CCPA's vague definition may also be a contributing factor. We found the following types of justifications, as shown in Table \ref{tab:sale}:

\begin{itemize}
    \item \textbf{Operational justification} (Excerpt \ref{quote:spotify}, \ref{quote:hulu}): The text highlights specific commercial activities or purposes that fall under the definition of a ``sale'', regardless of whether the business engages in those activities or purposes.
    
    \item \textbf{Legal justification} (Excerpt \ref{quote:cnn}, \ref{quote:twitch}): The text only refers to CCPA's definition of a ``sale'' and does not enumerate the types of commercial activities that may fall under it.
    
    \item \textbf{No justification} (Excerpt \ref{quote:healthline}): The text does not elaborate upon the stated stance in either of the preceding ways. This applies only to policies which state that the business does or does not sell personal information without qualification.
\end{itemize}

A majority (54.65\%) of sale disclosures state that the associated businesses do not sell personal information, and nearly half of them do not provide further elaboration. But do consumers have any reason to trust such statements, especially when similar businesses state that they \emph{do} ``sell'' personal information? As also shown in Table \ref{tab:sale}, each interpretation of the term ``sale'' is associated with explicit and implicit disclosures of advertising practices:

\begin{itemize}
    \item \textbf{Direct disclosure of advertisers} (Excerpt \ref{quote:salesforce}, \ref{quote:hulu}): The text lists advertising or marketing companies among the recipients of personal information shared by the business.
    
    \item \textbf{Indirect disclosure of advertising} (Excerpt \ref{quote:quizlet}): The text references targeted or personalized advertising or marketing as a purpose for which personal information is shared.
\end{itemize}

Such practices are described by all of the disclosures that assert it is sold in some way, but also by 72.34\% of disclosures that assert information is not sold. Thus, despite the variance in interpretations of the word ``sale'', at least some of these businesses are conceivably performing the same practices in reality. These patterns align with those found by \cite{OConnor2021} in their usability study of CCPA sale opt-outs. 

Especially considering the revelations of the Cambridge Analytica scandal and their wide-reaching impacts on consumer attitudes towards data privacy \cite{Berghel2018}, modern consumers negatively view the ``selling'' of data between businesses in scenarios where they lack the agency to control such practices \cite{Shipman2020}. As potential responses to either consumers or regulators, 84.88\% of disclosures thus state that no personal information is sold or attempt to qualify their ``sale'' of personal information. Yet, the common presence of advertising practices suggests that these statements may not translate into practice. We should thus question whether consumers can use the text of privacy policies to meaningfully differentiate between sharing practices that may or may not involve ``valuable consideration''. 

Our results support prior criticism of the term ``sale'' in CCPA, but they also document a new phenomenon of vagueness. This is because the distinction between ``sharing'' and ``selling'' is a novelty of CCPA; GDPR generically refers to the ``processing'' of personal information \cite{Lee2020}. Still, the practical issues caused by the vagueness of ``selling'' in CCPA mirror those caused by the term ``profiling'' in GDPR, which is defined as ``any form of automated processing of personal data consisting of the use of personal data to evaluate certain personal aspects relating to a natural person'' (\S 4.4). While GDPR provides the right for individuals to not be subject to ``solely automated decision-making'' (\S 22.1), it is similarly unclear what types of activities would fall under this definition. \cite{Gonzalez2019,Veale2018}

% CCPA's distinction would mitigate a case documented by \cite{Mohan2019}, where a user seeking to opt out of the use of their data for ``marketing and promotional purposes'' under GDPR was directed to delete their account. 

For privacy policies which state that no personal information is sold, there are four types of statements regarding whether consumers can still opt out of sales. As shown in Table \ref{tab:sale}, these are:

\begin{itemize}
    \item \textbf{Provision of opt-out} (Excerpt \ref{quote:healthline}): The text specifies a mechanism by which consumers can opt out of sales.
    
    \item \textbf{Hypothetical assurance of opt-out} (Excerpt \ref{quote:medium}): The text states that, if the business sells personal information in the future, the consumer will have an option to opt out.
    
    \item \textbf{Denial of opt-out} (Excerpt \ref{quote:linkedin}): The text states that the business does not offer an opt-out for the sale of personal information because it does not sell personal information.
    
    \item \textbf{Silence}: The text does not inform consumers about the right to opt out of the sale of personal information under CCPA.
\end{itemize}

CCPA regulations require the right to opt out of sale to be disclosed in privacy policies regardless of whether personal information is sold. Yet, a majority (76.6\%) of disclosures which state that personal information is not sold fall into the last type, which is potentially non-compliant. This contrasts with the setting of GDPR, where \cite{Zaeem2020} found that 90\% of privacy policies notified consumers of opt-outs to the \emph{processing} of personal information for marketing. The difference may involve the unpalatability of the word ``sale''.

Lastly, among the privacy policies which state that personal information is ``sold'' in some way, there are also varying levels of specificity about the categories of personal information involved:

\begin{itemize}
    \item \textbf{Full categorization} (Excerpt \ref{quote:cnn}): The text exhaustively lists all categories of personal information that may be sold.
    
    \item \textbf{Partial categorization} (Excerpt \ref{quote:indeed}): The text indicates that its listing of categories of sold personal information is incomplete (by ``for example'', ``such as'', or similar).
    
    \item \textbf{No categorization} (Excerpt \ref{quote:healthline}): The text generally refers to personal information in the context of selling information.
\end{itemize}

Helpfully, many (62.07\%) of the disclosures in this category do provide a comprehensive categorization, usually based on the categorization that CCPA mandates for right-to-know requests.

\subsection{Right to deletion}
\label{sec:policies:deletion}
\begin{table*}[ht]
\aboverulesep=0ex
\belowrulesep=0ex
    \centering
    \begin{tabular}{@{}c|c|c|cc|cs|c@{}}
        \toprule
        \rowcolor[rgb]{1,1,1}
        & & \multicolumn{5}{c|}{\textbf{Implementation of verification}} & \multirow{3}{*}{\textbf{Total}} \\
        & & \multirow{2}{*}{\textit{Info. request}} & \multicolumn{2}{c|}{\textit{Login wall}} & & & \\
        & & & \textit{Soft} & \textit{Hard} & \multirow{-2}{*}{\textit{Generic description}} & \multirow{-2}{*}{\textit{Silence}} \\
        \midrule
         & \textit{Present} & N/A & 5 (6.02\%) & 2 (2.41\%) & N/A & N/A & 7 (8.43\%) \\
        \rowcolor[rgb]{0.9,0.9,0.9} \cellcolor[rgb]{1,1,1}\multirow{-2}{*}{\textbf{Login wall explanation}} & \textit{Absent} & N/A & 31 (37.35\%) & 7 (8.43\%) & N/A & N/A & 38 (45.78\%) \\
        \midrule
        \multirow{2}{*}{\textbf{Retention --- basis}} & \textit{Specific} & 6 (7.23\%) & 12 (14.46\%) & 2 (2.41\%) & 5 (6.02\%) & 6 (7.23\%) & 19 (22.89\%) \\
        & \textit{Generic} & 8 (9.64\%) & 4 (4.82\%) & 2 (2.41\%) & 3 (3.61\%) & 4 (4.82\%) & 16 (19.28\%) \\
        \midrule
        \multirow{3}{*}{\textbf{Retention --- scope}} & \textit{Full} & 0 (0\%) & 2 (2.41\%) & 1 (1.2\%) & 2 (2.41\%) & 2 (2.41\%) & 4 (4.82\%) \\
        & \textit{Partial} & 0 (0\%) & 5 (6.02\%) & 1 (1.2\%) & 2 (2.41\%) & 3 (3.61\%) & 8 (9.64\%) \\
        & \textit{Generic} & 14 (16.87\%) & 9 (10.84\%) & 2 (2.41\%) & 4 (4.82\%) & 5 (6.02\%) & 23 (27.71\%) \\
        \midrule
        \multicolumn{2}{c|}{\textbf{No retention disclosed}} & 21 (25.3\%) & 20 (24.1\%) & 5 (6.02\%) & 6 (7.23\%) & 15 (18.07\%) & 48 (57.83\%) \\
        \midrule
        \multirow{2}{*}{\textbf{Propagation to third parties}} & \textit{Present} & 4 (4.82\%) & 3 (3.61\%) & 0 (0\%) & 0 (0\%) & 3 (3.61\%) & 7 (8.43\%) \\
        & \textit{Absent} & 31 (37.35\%) & 33 (39.76\%) & 9 (10.84\%) & 14 (16.87\%) & 22 (26.51\%) & 76 (91.57\%) \\
        \midrule
        \multirow{3}{*}{\textbf{Disruption of experience}} & \textit{Account-level} & 5 (6.02\%) & 10 (12.05\%) & 3 (3.61\%) & 3 (3.61\%) & 5 (6.02\%) & 16 (19.28\%) \\
        & \textit{Function-level} & 4 (4.82\%) & 5 (6.02\%) & 1 (1.2\%) & 1 (1.2\%) & 3 (3.61\%) & 8 (9.64\%) \\
        & \textit{None} & 26 (31.33\%) & 21 (25.3\%) & 5 (6.02\%) & 10 (12.05\%) & 17 (20.48\%) & 59 (71.08\%) \\
        \midrule
        \multicolumn{2}{c|}{\textbf{Total}} & 35 (42.17\%) & 36 (43.37\%) & 9 (10.84\%) & 14 (16.87\%) & 25 (30.12\%) & 83 \\
        \bottomrule
    \end{tabular}
    \medskip
    \caption{Summary of patterns in deletion disclosures across all 83 unique privacy policies with instructions for deletion requests. See text of Section \ref{sec:policies:deletion} for pattern definitions. Percentages show the proportion of all policies with instructions for deletion requests that exhibit a given pattern or combination of patterns. Shaded cells denote possible non-compliance.}
    \label{tab:deletion}
\vspace{-2em}
\end{table*}

All but three privacy policies (which lack CCPA-specific disclosures) specify at least one method by which consumers can submit deletion requests. Beyond this, however, substantial variation exists. As shown in Table \ref{tab:deletion}, we observed the following patterns concerning the description of the process for verifying deletion requests:

\begin{itemize}
    \item \textbf{Information request} (Excerpt \ref{quote:ebay}): The text states that the business may request additional information to verify the consumer's deletion request before processing it.
    
    \item \textbf{Login wall} (Excerpt \ref{quote:pinterest}): The text provides different steps for submitting deletion requests to users who do and users who do not hold accounts with the business.
    
    \begin{itemize}
        \item \textbf{Hard login wall} (Excerpt \ref{quote:netflix}): The text only provides instructions for submitting deletion requests to users who hold accounts, and not to non-account holders.
    \end{itemize}
    
    \item \textbf{Generic description} (Excerpt \ref{quote:wp}): The text states that the business will verify the consumer's identity before processing the deletion request, but does not specify what actions the consumer needs to take after submitting their request.
    
    \item \textbf{Silence}: The text does not discuss verification.
\end{itemize}
        
% \item \textbf{Hard login wall} (Excerpt \ref{quote:netflix}): The text clearly states that deletion requests may only be submitted by users who hold accounts, and not non-account holders.

Given the latitude in what ``general descriptions'' of verification processes may consist of, all but the last of these patterns would likely be compliant with CCPA regulations (\S 999.308 (c)(2)). Yet, a substantial minority (30.12\%) of disclosures are indeed silent about their verification processes. Silence may imply that \emph{no verification} takes place, which could be ``reasonable'' if unauthorized access or deletion would not affect consumers (\textit{cf.} \S 999.323 (b)(3)b). 

As for ``login walls'', such distinctions can be reasonably expected for types of personal information that will only be collected from account holders. They in fact conform to CCPA regulations, which require businesses to verify the identity of non-account holders ``to a reasonable degree of certainty'' (\S 999.325 (b)), and to deny deletion requests to users for which this is impossible (\S 999.325 (f)). However, the latter must be clearly explained in privacy policies (\S 999.325 (g)), which we only observed in 15.56\% of these policies:

\begin{itemize}
    \item \textbf{Login wall explanation} (Excerpt \ref{quote:netflix}): The text states that deletion requests from users who lack accounts may be rejected because they cannot be verified to CCPA's standards.
\end{itemize}

Businesses also provide varying levels of detail about the possible retention of personal information after a successful deletion request, as also summarized in Table \ref{tab:deletion}. The purposes for retaining personal information may be described in one of the following ways:

\begin{itemize}
    \item \textbf{Specific operational basis} (Excerpt \ref{quote:wp}): The text makes reference to specific business purposes that are exempt under CCPA to explain why information may be retained.
        
    \item \textbf{Generic operational basis} (Excerpt \ref{quote:co}): The text does not specify which types of commercial activities personal information would need to be retained for, including when the text lists all exempt business purposes under CCPA.
\end{itemize}

Meanwhile, the categories of personal information that are retained may be described in one of the following ways:
\begin{itemize}
    \item \textbf{Full categorization}: The text exhaustively lists all categories of personal information that may be retained.

    \item \textbf{Partial categorization}: The text indicates that its listing of categories of personal information that may be retained is incomplete (by ``for example'', ``such as'', or similar).
    
    \item \textbf{Generic categorization}: The text generally refers to personal information that may be retained.
\end{itemize}

Notably, the 42.17\% of privacy policies which state that personal information may be retained all specify at least one operational basis for which this information may be retained, and 54.29\% of such disclosures are reasonably specific. Meanwhile, only 34.29\% of them describe the categories of information that may be retained. CCPA regulations do not require the latter disclosures for privacy policies; their presence may stem from GDPR disclosure requirements regarding retention, particularly \S 13.2(a) and \S 14.2(a) \cite{Linden2020}. These inconsistencies in the clarity of retention purposes and scopes echo the findings of \cite{Contissa2018,Linden2020,Mohan2019,Zaeem2020} within the context of GDPR. 

These vague statements of retention purposes also point to a pitfall in CCPA's definition of ``business purposes'': we believe that businesses could share or sell the retained personal information depending on their interpretation of ``reasonably anticipated'' commercial activities. Only four privacy policies state otherwise:

\begin{itemize}
    \item \textbf{Propagation to third parties} (Excerpt \ref{quote:walmart}): The text states that the business will relay deletion requests to other parties with access to the consumer's personal information (such as service providers or clients).
\end{itemize}

While this procedure is mandated by CCPA (\S 1798.105(c)), few consumers will probably be aware of this provision (indeed, few consumers are aware of CCPA's provisions; see Section \ref{sec:survey-results}). Without seeing such disclosures, consumers would likely underestimate the extent of their privacy's protection under CCPA (see Section \ref{sec:survey-results:deletion}).

Lastly, a sizeable minority (28.92\%) of deletion disclosures state that a user's experience with the business may change after their personal information is deleted. We observed two such types of patterns, as also summarized in Table \ref{tab:deletion}:

\begin{itemize}
    \item \textbf{Account-level disruption} (Excerpt \ref{quote:bb}): The text states that a successful deletion request could result in the deletion or deactivation of the user's entire account.

    \item \textbf{Function-level disruption} (Excerpt \ref{quote:wp}, \ref{quote:zillow}): The text states that a successful deletion request may hinder the user's experience in other ways, including through the loss of access to functionality requiring personal information. 
\end{itemize}

Both the privacy policies themselves and CCPA are sufficiently vague that these practices may fall under the exemption in the non-discrimination clause for differences that are ``reasonably related'' to the value of the personal information (\S 1798.125(b)). Still, the deactivation of user accounts in the former case is a striking instance of the ``take it or leave it'' approach to privacy controls, which has also been criticized in the context of GDPR \cite{Contissa2018,Gonzalez2019,Mohan2019}. 

In summary, beyond the vague requirements of CCPA regulations, the privacy policies that we analyzed provide inconsistent levels of detail about the request verification process as well as the retention and use of personal information following successful deletion requests. These vague disclosures may disincentivize users from exercising their privacy rights because they reduce the perceived effectiveness of deletion requests.

\section{User survey background}
\label{sec:survey-bg}
To evaluate the effect of varying interpretations of CCPA's privacy policy requirements, we conducted an IRB-approved user survey of California residents to assess whether their understanding of businesses' data practices depends on the specificity of these businesses' CCPA-mandated privacy disclosures. We focused on three patterns of variance, each selected from one of the three types of disclosures pertaining to data flows that we focused on in our privacy policy analysis: the mapping between recipients and categories of shared personal information (found in sharing disclosures), the types of commercial activities affected by the right to opt out of sale (found in sale disclosures), and the effects of successful deletion requests on data retention and sharing (found in deletion disclosures). 

In our user survey, we asked respondents to read three excerpts (each containing one of the three types of disclosures) from the CCPA disclosures of a single privacy policy. Then, for each excerpt, we presented them with two data usage scenarios based on the content of the disclosure, and we asked the respondents to rate the likelihood that the business would be allowed or required to engage in each of the described data practices. We did not require respondents to rely on the policy alone; thus, they were free to draw upon any prior knowledge they had about relevant regulations. 

\subsection{Policy selection}
\label{sec:survey-bg:policy}
To maintain the practical feasibility of our study, we randomly and (as much as possible) evenly showed excerpts from one of nine privacy policies to the respondents, as summarized in Table \ref{tab:policies}. These policies were selected so that they contained substantial variance in the specificity of the disclosures that we considered.

% (1) Walmart; (2) eBay; (3) New York Times; (4) Hulu; (5) ESPN; (6) Best Buy; (7) Google; (8) Microsoft; and (9) Netflix

\begin{table*}[ht]
    \centering
    \begin{tabularx}{\textwidth}{@{}llXXXXXX@{}}
        \toprule
        \textbf{Privacy Policy} & $\boldsymbol{n}$ & \textbf{Sharing: \newline Recipient / Info Mapping} & \textbf{Sharing: \newline Sharing \newline Purposes} & \textbf{Sale: \newline Interpretation of ``Sale''} & \textbf{Sale: \newline Justification of Stance} & \textbf{Deletion: \newline Retention \newline Purposes} & \textbf{Deletion: \newline Propagation to 3rd Parties} \\
        \midrule
        \textbf{Section Ref} & & \ref{sec:policies:sharing} & \ref{sec:policies:sharing} & \ref{sec:policies:sale} & \ref{sec:policies:sale} & \ref{sec:policies:deletion} & \ref{sec:policies:deletion} \\
        \midrule
        (1) Walmart & 40 & One-to-one \newline mapping & Examples by \newline recipient & Sharing & Legal & Generic & Service providers \\
        (2) eBay & 41 & Separate lists & Generic list & Sharing & Operational & Generic & Not specified \\
        (3) New York Times & 41 & Examples by \newline recipient & Generic list & Sharing & Legal & Records / transactions & Not specified \\
        (4) Hulu & 41 & Examples by \newline recipient & Examples by \newline recipient & Advertising & Operational & Generic & Not specified \\
        (5) ESPN & 40 & Separate lists & Generic list & Advertising & Legal & Generic & Not specified \\
        (6) Best Buy & 40 & One-to-one \newline mapping & Generic list & Advertising & Operational & Transactions & Not specified \\
        (7) Google & 40 & No list of \newline shared PI & Generic list & No sale & Unjustified & Security / \newline records & Not specified \\
        (8) Microsoft & 40 & Mapping & Mapping by PI & No sale & Unjustified, \newline no opt out & Generic & Not specified \\
        (9) Netflix & 41 & One-to-one \newline mapping & Examples by \newline recipient & No sale & Unjustified & Records / \newline legal & Not specified \\
        \bottomrule
    \end{tabularx}
    \caption{Summary of the nine privacy policies used for our survey (Section \ref{sec:survey-bg:policy}). $n$ is the number of survey respondents.}
    \label{tab:policies}
\vspace{-2em}
\end{table*}

For sharing disclosures (Section \ref{sec:policies:sharing}), 4 of the policies we selected provide a one-to-one mapping between the categories of personal information that are shared by the business and the categories of recipients. Meanwhile, 2 policies provide examples of categories of personal information for each category of recipients, and 3 policies make no association between personal information and recipients. 

Next, for sale disclosures (Section \ref{sec:policies:sale}), 3 policies each describe (1) sharing activities in general as potentially being sales; (2) advertising activities as sales; and (3) no activities as sales. Both categories that disclose ``sales'' consist of policies that base their stances on CCPA's definition or on specific commercial purposes. Among the policies that do not disclose ``sales'', only one policy states that the consumer has no right to opt out of the sale of personal information. 

Lastly, for deletion disclosures (Section \ref{sec:policies:deletion}), 5 policies do not specify why information may be retained, 3 mention record-keeping as a purpose, and 2 mention transactions as a purpose. Only one specifies that deletion requests are relayed to service providers. 

Thus, for each type of disclosure, at least a third of the policies that we selected for our user survey are substantially vaguer than the others. Our power analysis uses this property (Appendix \ref{sec:appendix:power}).

\subsection{Research questions}
\label{sec:survey-bg:rq}
Here, we list the basic types of questions in our user survey; Appendix \ref{sec:appendix:survey} contains the full user survey, including the privacy policy excerpts. These questions were adjusted based on the excerpts, including different instantiations for the template fields. Separate versions of \textbf{Q3} and \textbf{Q4} were needed because (1) sharing information in exchange for advertising revenue falls under the definition of a ``sale'' in policies from the ``Advertising'' category and (2) sale opt-outs do not apply to policies from the ``No sale'' category.

\begin{enumerate}[\textbf{Q\arabic*}]
    \item[] \textbf{Sharing disclosures}

    \item If you make purchases through a website owned by Company XYZ, is Company XYZ allowed to share your \textbf{past purchase history} with a payment service provider? \medskip
    
    \item If you make purchases through a website owned by Company XYZ, is Company XYZ allowed to share your \textbf{inferred ethnicity} (inferred based on your purchase history) with a payment service provider? \medskip
    
    \newpage
    \item[] \textbf{Sale disclosures}

    \item 
    \textit{``Sharing''}
    \begin{itemize}
        \item If you opt out of the sale of your personal information, is Company XYZ allowed to provide your \textbf{full name and mailing address} to an advertising company, \textbf{without receiving revenue in return} from this advertising company? \medskip
    \end{itemize}
    
    \textit{``Advertising''}
    \begin{itemize}
        \item If you opt out of the sale of your personal information, is Company XYZ allowed to provide your \textbf{full name and mailing address} to \textbf{an advertising company} so that the advertising company can display advertisements to you on other websites, \textbf{without receiving revenue in return} from this advertising company? \medskip
    \end{itemize}
     
    \textit{``No sale''}
    \begin{itemize}
        \item Could Company XYZ be allowed to provide your \textbf{full name and mailing address} to an advertising company, \textbf{without receiving revenue in return} from this advertising company? \medskip
    \end{itemize}

    \item 
    \textit{``Sharing''}
    \begin{itemize}
        \item If you opt out of the sale of your personal information, is Company XYZ allowed to provide your \textbf{full name and mailing address} to an advertising company, \textbf{in return for a portion of revenue} from this advertising company? \phantom{}
    \end{itemize}
    
    \textit{``Advertising''}
    \begin{itemize}
        \item If you opt out of the sale of your personal information, is Company XYZ allowed to provide your full name and mailing address to \textbf{a separately-owned [category of Company XYZ]} so that the two companies can provide joint product offers to you, \textbf{without receiving revenue in return} from this partner company? \medskip
    \end{itemize}
     
    \textit{``No sale''}
    \begin{itemize}
        \item If Company XYZ provides your \textbf{full name and mailing address} to an advertising company, without receiving revenue in return from this advertising company, would you \textbf{have an option to opt-out} of this sharing? \medskip
    \end{itemize}
    
    \item[] \textbf{Deletion disclosures}

    \item If you request the deletion of your personal information, is Company XYZ required to direct \textbf{any marketing company that has received your information, either by sharing or selling, from Company XYZ} to delete information \\ about your activity on Company XYZ's websites? \smallskip
    
    \item If you request the deletion of your personal information, is Company XYZ required to direct \textbf{any marketing company that provides advertising consultation services to Company XYZ} (that is, acting in the capacity of a service provider) to delete information about your activity on Company XYZ's websites? \medskip
\end{enumerate}

For each question, we asked respondents to answer using a 5-point Likert scale of (1) ``Definitely not'', (2) ``Probably not'', (3) ``I'm really not sure'', (4) ``Probably'', and (5) ``Definitely''. We then developed a set of ground truth consensus answers from three of the authors who are experts in privacy law and practice. These authors first individually answered all six questions corresponding to each of the nine policies (54 total) using the same Likert scale. Then, all disagreements with the majority answer were discussed and resolved based on the text of CCPA and CCPA regulations. 

To allow for uncertainty in the respondents' answers, we marked them as correct as long as they were sufficiently close to our consensus. For a ground truth of 4 or 5, we accepted both 4 and 5; for 3, we accepted 2, 3, or 4; and for 1 or 2, we accepted both 1 and 2. Based on our evaluation of the correctness of the respondents' answers, we formulated the following research question:
\begin{itemize}[\textbf{RQ}]
    \item For each question (\textbf{Q1}, \textbf{Q2}, \textbf{Q3}, \textbf{Q4}, \textbf{Q5}, and \textbf{Q6}), is the proportion of respondents who answer correctly independent of the privacy policy excerpt that is shown to them?
\end{itemize}

\subsection{Survey recruitment} 
\label{sec:survey-bg:recruitment}
\textbf{RQ} leads to a $\chi^2$ test of independence for each question between two categorical factors: (1) the privacy policy that is shown and (2) answer correctness. For \textbf{Q1}, \textbf{Q2}, \textbf{Q3}, \textbf{Q5}, and \textbf{Q6}, the privacy policy factor has 9 levels. We conduct three tests for \textbf{Q4} and three additional tests for \textbf{Q3} because the questions for the three types of opt out disclosures are not entirely comparable; the privacy policy factor for each of these tests has 3 levels. Our power analysis (see Appendix \ref{sec:appendix:power}) led us to recruit a sample of 364 respondents.

Our sample of respondents was recruited through Amazon Mechanical Turk. Respondents were required to be California residents who were at least 18 years old. To avoid low-quality responses, we sampled respondents from the pool of Mechanical Turk users with an approval rate of at least 90\%. Before participating in the study, respondents were required to complete a screening questionnaire and a consent form; any respondents who failed to certify that they met the criteria were removed from the participant pool. 

\subsection{Survey methodology}
\label{sec:survey-bg:methods}
Respondents were directed to complete (1) a pre-survey questionnaire, (2) the main survey, and (3) a post-survey questionnaire. 

\begin{enumerate}[(1)]
    \item We first assessed respondents' familiarity with privacy legislation by asking them to rate their knowledge of CCPA, CPRA, CalOPPA, and GDPR.
    \smallskip
    
    \item In the main survey, we implemented a between-subjects design where each respondent was shown three excerpts from a single privacy policy in a randomized order. All company names, website URLs, and telephone numbers were anonymized to prevent priming effects. We reproduced the original formatting of the text as closely as possible, except we removed all hyperlinks. These methods are consistent with \cite{Vu2007,Vail2008,Korunovska2020} but not \cite{Reeder2008} (who compared alternate representations of the same text using a within-subjects design).
    % Specifically, we changed all company names to ``Company XYZ''. 
    
    Note that we only displayed excerpts instead of the entire privacy policy. This is because the goal of our user survey was to assess the ability of consumers to understand and interpret the contents of the excerpts. Indeed, the orthogonal issue of whether consumers can \emph{locate} relevant text in a privacy policy has already been studied by \cite{Vu2007}. Our goals and thus our methodology are closer to \cite{Tang2021} (who assessed comprehension of specific terms in the context of sample sentences) and \cite{Bui2021}. After each excerpt, we showed each respondent the two questions corresponding to the excerpt on the same page (the questions listed in Section \ref{sec:survey-bg:rq} were worded to avoid priming effects). Again, this was because we aimed to assess comprehension instead of memorization. This approach differs from \cite{Vu2007,Vail2008} but aligns with \cite{Korunovska2020}.
    
    For each question, we provided two free-text response fields where we asked respondents (1) to paste one to two sentences from the excerpt that most influenced their answer, and (2) to briefly explain their answer. We evaluated the effort level of respondents using their free-text responses, along with two simple multiple-choice attention checks based on the content of the questions. We withdrew respondents who provided irrelevant answers for both types of questions.
    \smallskip
    
    \item Afterward, we presented respondents with two sets of post-survey questions, namely a demographic questionnaire and a standard privacy/security questionnaire \cite{Faklaris2019}.
\end{enumerate}

In a pilot of 15 respondents, the mean completion time was 16 minutes, 52 seconds. Using an hourly rate of \$12.00, we compensated respondents with a nominal base amount of \$3.00. To incentivize respondents to consider their answers carefully, we provided a bonus of \$0.15 for every correct answer, plus another \$1.00 if they correctly answered all six questions; we provided a mean bonus of \$0.55. The median response time for all respondents was 20 minutes, 10 seconds, with the first quartile being 14 minutes, 52 seconds, and the third quartile being 27 minutes, 51 seconds.

We compared the self-reported demographics of our respondents to California's overall demographics from the 2019 American Community Survey \cite{ACS2019}. Our sample slightly under-represents males (49.4\% compared to 49.7\%); over-represents single people (58.45\% compared to 51.6\%), and over-represents people with a college degree or higher (74.45\% compared to 35\%). Respondents' knowledge of privacy regulations was variable; 56.9\% had heard of CCPA, versus 39.6\% for CPRA, 25.8\% for CalOPPA, and 23.6\% for GDPR.

\section{User survey results}
\label{sec:survey-results}
\begin{figure*}[ht]
\centering
\begin{subfigure}{0.3\textwidth}
    \centering
    \includegraphics[width=\textwidth]{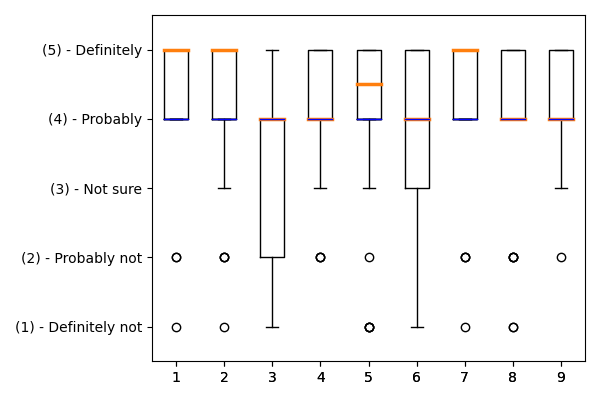}
    \caption{\textbf{Q1}: Sharing, purchase history}
    \label{fig:bp_q1}
    \includegraphics[width=\textwidth]{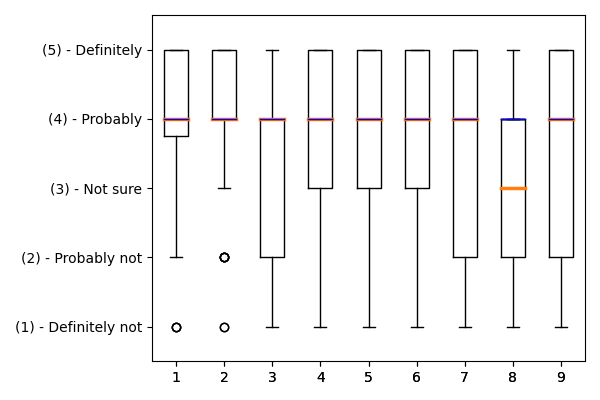}
    \caption{\textbf{Q2}: Sharing, inferred ethnicity}
    \label{fig:bp_q2}
\end{subfigure}
\hfill
\begin{subfigure}{0.3\textwidth}
    \centering
    \includegraphics[width=\textwidth]{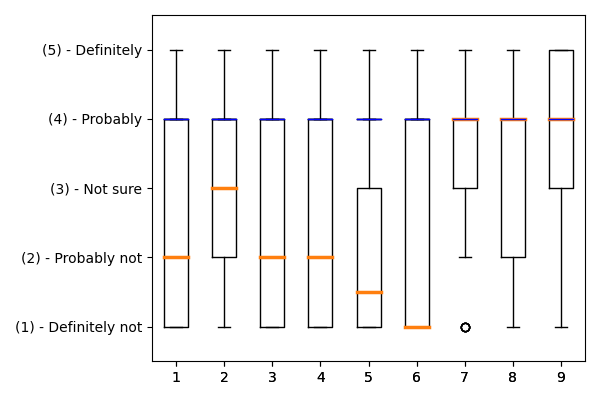}
    \caption{\textbf{Q3}: Sale, without revenue exchange}
    \label{fig:bp_q3}
    \includegraphics[width=\textwidth]{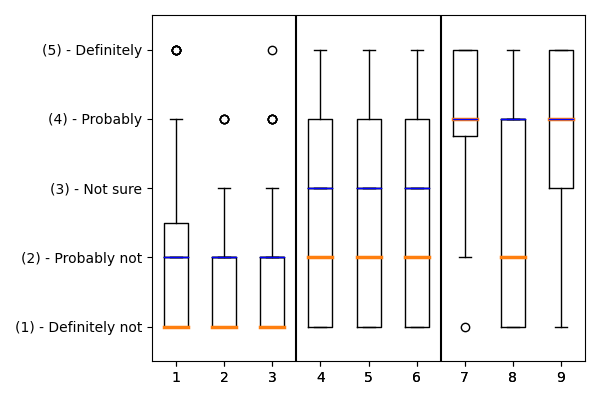}
    \caption{\textbf{Q4}: Sale, revenue / Joint offer / Opt-out}
    \label{fig:bp_q4}
\end{subfigure}
\hfill
\begin{subfigure}{0.3\textwidth}
    \centering
    \includegraphics[width=\textwidth]{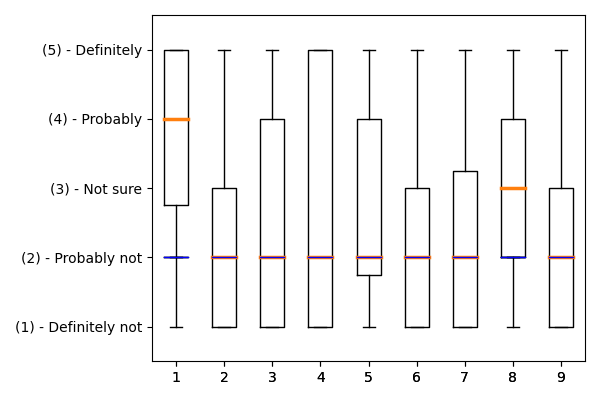}
    \caption{\textbf{Q5}: Deletion, all third parties}
    \label{fig:bp_q5}
    \includegraphics[width=\textwidth]{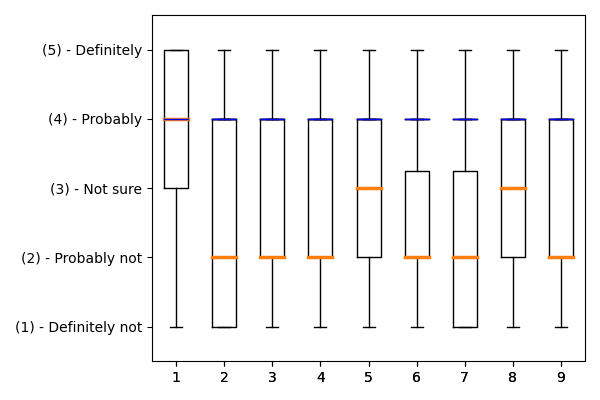}
    \caption{\textbf{Q6}: Deletion, service providers}
    \label{fig:bp_q6}
\end{subfigure}
\caption{Boxplots of Likert scale survey responses from 364 respondents. Orange lines show the median respondent answer; blue lines show the ground truth. Numbers on the horizontal axis correspond to the privacy policies listed in Section \ref{sec:survey-bg:policy}.}
\label{fig:bp}
\end{figure*}

In Figure \ref{fig:bp}, we show the distribution of answers for each question and policy. Our overall results in terms of question correctness and our statistical tests are summarized in Table \ref{tab:chisq} in Appendix \ref{sec:appendix:correct-chisq}. 

Based on their free-text responses, we found that respondents rarely referenced legal knowledge. Only 5.52\% of rationales explicitly mentioned the definitions or provisions of CCPA or other California laws. This was even true for respondents who self-reported that they had heard of CCPA (6.18\% of rationales from these respondents mentioned CCPA), which suggests that an individual knowing of CCPA is not a strong signal that they will use that knowledge to evaluate their privacy in practice. At the same time, 28.18\% of rationales mentioned that the excerpts alone did not address the scenarios clearly enough for them to give a certain answer.

\subsection{Sharing disclosures}
\label{sec:survey-results:sharing}
For \textbf{Q1} and \textbf{Q2}, our ground truth states that both purchase history and inferred ethnicity could probably be shared by all of the businesses, since they fall under the categories of personal information that could be disclosed to service providers according to these businesses' privacy policy excerpts. We found that 80.77\% of respondents agreed with our consensus answer for \textbf{Q1}, but only 63.74\% of respondents agreed with our consensus answer for \textbf{Q2}.

Our statistical tests indicated that there was a statistically significant difference in answer correctness between the nine policies on both \textbf{Q1} ($p \approx 0.0113$) and \textbf{Q2} ($p \approx 0.0321$). For all privacy policies except those for New York Times, Hulu, ESPN, and Best Buy, 4 (``Probably allowed'') was the first quartile of answers to \textbf{Q1}, but this was only true of Walmart and eBay on \textbf{Q2}. We infer that respondents were overall more certain that sharing would be permitted for \textbf{Q1} but were less confident on some policies, while they were more ambivalent overall for \textbf{Q2}.

In 87.28\% of rationales for correct answers on both questions (83.38\% of all rationales), respondents referenced list entries that contain relevant categories of shared personal information and recipients. This was even true for policies that provide no association between them, on which respondents inferred the sharing of \emph{every} category of personal information with \emph{every} category of recipients. 

However, many respondents were confused about the correct categorizations that the presented scenarios fell under. We found that 35.53\% of rationales for incorrect answers (22.63\% overall) did not identify the correct category of personal information in the disclosure. For \textbf{Q2} in particular, 40.34\% and 29.97\% of all rationales considered inferred ethnicity as a demographic or as an inference (both of which are plausible), and 9.52\% incorrectly classified it as an identifier. Meanwhile, 14.72\% (16.9\% overall) of rationales showed confusion about recipients (including the conflation of ``service providers'' and ``third parties'', which are distinct under CCPA).

\subsection{Sale disclosures}
\label{sec:survey-results:sale}
For \textbf{Q3}, our ground truth states that businesses would probably be allowed to share information without receiving revenue in return, provided that they interpret these activities as not falling under CCPA's definition of a ``sale'' and any benefits they receive as not representing ``valuable consideration''. Overall, respondents struggled with these definitions. For policies in the ``Sharing'' and ``Advertising'' categories, the majority of respondents disagreed with our answer, with only 38.52\% and 30.58\% of their answers respectively matching our consensus. Among incorrect answers, 38.85\% of rationales considered this revenue-less scenario to fall under the opt-out for sales, while 28.66\% interpreted the opt-out as covering all forms of sharing. Meanwhile, the proportion of correct answers for ``No sale'' was much higher, at 65.29\%. For correct answers, 68.35\% of these rationales \emph{excluded} non-monetary forms of sharing from the definition of ``sale'' --- explicit statements from the businesses that they do not sell personal information appear to have reduced respondents' perceived scope of a ``sale''. The differences across all policies are statistically significant ($p \approx 1.944 \times 10^{-6}$), although the question is worded differently for the three categories. Within-category statistical tests indicated that only the ``Sharing'' category had marginally significant inter-policy differences ($p \approx 0.0488$).

On \textbf{Q4} for the ``Sharing'' policies, 80.33\% of respondents agreed with our ground truth that exchanging personal information for ad revenue would likely constitute a ``sale'', and thus would probably not be allowed after they have opted out of the sale of personal information. Indeed, 60.2\% of rationales for correct answers recognized that the scenario fell under the opt-out. However, 33.33\% of rationales for incorrect answers emphasized the ambiguity in the definition of a ``sale''. The statistical tests did not find significant differences in correctness between the policies in this category.

For the ``Advertising'' policies, our ground truth recognizes that the answer to \textbf{Q4} would be dependent on how regulators interpret the term ``valuable consideration''. Partner businesses in a co-marketing relationship might not exchange any revenue from their sales. Even so, joint offers would likely have super-additive benefits to the customer bases of both businesses, which may be ``valuable''. CCPA is much vaguer in this regard than the Gramm-Leach-Bliley Act, which applies to financial institutions and provides exemptions for co-marketing \cite{Cranor2016}. Despite this ambiguity, we found that 54.55\% of respondents were confident enough to give a definite answer. Again, we found no statistically significant inter-policy differences.

Lastly, on \textbf{Q4} for the ``No sale'' policies, our ground truth considers that third party advertisers who receive information from the businesses would likely provide opt-outs for such sharing practices, regardless of whether the first party provides an opt-out. While 61.16\% of respondents answered the question correctly, their reasoning was often incorrect. We found that 52.05\% of rationales for correct answers identified the right to deletion as the right to opt out; only 24.66\% mentioned an opt-out separately from deletion. Our statistical test found significant differences in correctness between policies ($p \approx 0.0008$); this was likely related to Microsoft's statement that they do not provide an opt-out, which was mentioned in 83.33\% of rationales for incorrect answers on this policy.

\subsection{Deletion disclosures}
\label{sec:survey-results:deletion}
For \textbf{Q5}, according to our ground truth, CCPA requires businesses to relay deletion requests to service providers, but probably not to all third party recipients. Overall, 57.69\% of respondents agreed with our conclusion. For correct answers, 66.67\% of rationales recognized that the right to deletion does not cover all third parties, and 47.34\% noted that the text does not discuss third parties. Our statistical test showed significant differences in correctness between policies ($p \approx 0.0010$); the median answers for two policies differ from our ground truth. Walmart's policy states that they would ``direct any service providers'' to carry out deletion requests; 70\% of incorrect rationales interpreted this as applying to third parties. The brevity of Microsoft's description likely led 45.45\% of incorrect rationales to suggest that the excerpt did not clearly address the scenario. 

Lastly, for \textbf{Q6}, our ground truth states that businesses would probably be required to relay deletion requests to marketing companies acting as service providers, again based on CCPA. Only 35.44\% of respondents arrived at the same answer; the median answer was unchanged from \textbf{Q5} on six out of nine policies. For incorrect answers, 70\% of rationales (often correctly) recognized that the policies did not address service providers either. Yet, this was also the case for 42.85\% of such rationales on Walmart, despite the policy's clear statement. They were confused by the service provider being a marketing company; such confusion may be alleviated by reiterating examples of service providers in the text. We again found significant inter-policy differences in correctness ($p \approx 0.0010$). 

% The median answer decreased for Microsoft, as several respondents' answers changed from ``Not sure'' to ``Probably not''; their rationales do not elucidate their reasoning. For respondents with correct answers, however, 32.28\% of rationales identified that special provisions likely apply to service providers. 

In summary, deletion disclosures led many respondents to form an incorrect but less harmful conclusion: that they would not be protected, whereas in reality they would be. However, as we suggested in Section \ref{sec:policies:deletion}, this may also reduce their confidence in deletion requests. Regardless, our findings highlight a lack of awareness regarding CCPA's deletion requirements for service providers.

\section{Conclusion}
\label{sec:conclusion}
In this work, we evaluated the clarity and effectiveness of currently implemented CCPA disclosures at informing consumers of their privacy rights. From our analysis of 95 privacy policies on popular websites, we found that the level of detail in disclosures was often inconsistent, with many of them obfuscating important information that would be relevant to consumers' privacy concerns. We also suggest that not all disclosures describe businesses' data practices at the level of clarity mandated by CCPA regulations. Our user survey then showed that different types of wording led consumers to different conclusions about businesses' data practices and their own privacy rights, even when the legal ground truth was the same.

\subsection{Limitations}
\label{sec:conclusion:limits}
Due to constraints of practicality, we were unable to perform a more comprehensive evaluation of CCPA disclosures through either our privacy policy analysis or our user survey. Our privacy policy analysis was restricted to 95 of the most popular websites across the entire United States. The level of variance among disclosures may differ for less popular websites; preliminary analysis suggests that boilerplate text is more common. However, the privacy policies we analyzed are likely representative of those on most websites that the average California resident would visit (see Section \ref{sec:policies}). We did not try to expand our analysis using natural language processing due to the error inherent in such approaches (\textit{cf.} \cite{Ammar2012,Sadeh2013,Zimmeck2017,Wilson2018,Linden2020}), which would hinder an accurate assessment of the status quo of privacy policy clarity under CCPA. Nevertheless, automating the analysis of CCPA disclosures is a valuable direction of future work that could improve their interpretability by consumers and policy-makers.

Our user survey also considered only a subset of the patterns found in CCPA disclosures. Although we included a diverse sample of excerpts, we could not capture the entire space of variation due to the highly customized nature of CCPA disclosures on popular websites. Still, the statistically significant effects that we observed, in spite of our limited sets of policies and respondents, suggest that variation in wording does impact consumers' perceptions. Also, our experimental design attempted to mitigate brand-based priming effects (\textit{cf.} \cite{Laran2010}) as much as possible. Based on the respondents' rationales, we do not believe that any of them made associations with real businesses, but we cannot entirely exclude this possibility.

\subsection{Implications for policy makers}
\label{sec:conclusion:policy}
Our results echo prior literature in highlighting problematic ambiguities in CCPA's terminology. Specifically, our privacy policy analysis showed that businesses do not consistently interpret the terms ``sale'', ``valuable consideration'', or ``business purpose'', nor CCPA regulations' requirements for privacy policies. These effects translated to confusion among survey respondents because the policy excerpts often failed to clarify the businesses' assumptions.

CPRA, which is intended to address CCPA's flaws, will soon come into force. While CPRA addresses some of our findings, it may introduce more ambiguity in other areas. CPRA clarifies that non-personalized advertising is a business purpose; provides an opt-out for targeted advertising; extends the requirement for relaying deletion requests to all third parties, not just service providers; and introduces a GDPR-like data minimization requirement \cite{CPRA2020,NAI2020}. Yet, CPRA also distinguishes ``contractors'' from service providers as entities to which businesses merely \emph{make personal information available}, which has already been considered unclear \cite{NAI2020}. 

Many states, such as Virginia, are now following suit in passing CCPA-like laws \cite{Kaminski2020}. If these laws are to succeed, policy makers and regulators must exercise due diligence in addressing vagueness and ambiguity through not just legislation but also enforcement, as suggested by the apparent non-compliance we observed. Encouragingly, some of these laws provide more clarity than CCPA \cite{Rippy2021} (\textit{e.g.}, the monetary definition of a ``sale'' in Nevada's new privacy law). Still, we argue that more standardized and structured disclosures based on a uniform interpretation of terminology would better mitigate the issue of vague and ambiguous privacy policies.

\section*{Acknowledgments}
This research was supported in part by grants from the National Science Foundation Secure and Trustworthy Computing program (CNS-1801316, CNS-1914486, CNS-1914444, CNS-1914446), as well as NSF grants IIS-1850477 and IIS-2046640 (CAREER). Co-author Rex Chen is supported in part by a Yahoo InMind Fellowship.

\bibliographystyle{plain}

\appendix
\appendixpage
\aboverulesep=0ex
\belowrulesep=0ex
\section{Privacy policy excerpts}
\label{sec:appendix:quotes}

\begin{myquote}{Microsoft privacy policy}
\begin{itemize}[leftmargin=-0.5em]
    \item \textbf{Name and contact data}
    \begin{itemize}
        \item Sources of personal data: Interactions with users and partners with whom we offer co-branded services
        \item Purposes of Processing (Collection and Sharing with Third Parties): Provide our products; respond to customer questions; help, secure, and \\ troubleshoot; and marketing
        \item Recipients: Service providers and user-directed \\ entities
    \end{itemize}
\end{itemize}
\label{quote:microsoft}
\end{myquote}

\begin{myquote}{Amazon privacy policy}
The personal information that Amazon disclosed to [third parties] for a business purpose in the twelve months prior to the effective date of this Disclosure fall into the following categories established by the California Consumer Privacy Act, depending on which Amazon Service is used:
\begin{itemize}
    \item identifiers such as your name, address, phone numbers, or IP address, for example if we use a third party carrier to deliver your order [...]
\end{itemize}
\label{quote:amazon}
\end{myquote}

\begin{myquote}{New York Times privacy policy}
\textbf{With Whom Do We Share the Information We Gather?} \\
\textbf{C) With Other Third Parties:}
\begin{itemize}[i]
\item If you’re a U.S. print subscriber, we may share your name and mailing address (among other information) with other reputable companies that want to market to you by mail.
\end{itemize}
\label{quote:nyt}
\end{myquote}

\begin{myquote}{Instructure privacy policy}
\begin{tabularx}{0.81\columnwidth}{@{}|X|X|@{}}
    \toprule
    \textbf{Categories of Personal Information} & \textbf{Disclosed for business purposes to the following categories of third parties:} \\
    \midrule
    Personal and online identifiers (such as first and last name, email address, or unique online identifiers) & All categories listed below \\
    \bottomrule
\end{tabularx}
\label{quote:instructure}
\end{myquote}

\begin{myquote}{Fedex privacy policy}
We collect, share and disclose Personal Information for the business and commercial purposes described in the "Why does FedEx process Personal Data?" and "Who has access to your Personal Data?" sections above. 
\label{quote:fedex}
\end{myquote}

\begin{myquote}{Salesforce privacy policy}
We may share your Personal Data as follows:
\begin{itemize}
    \item Service Providers: With our contracted service providers, who provide services such as IT and system administration and hosting, credit card processing, research and analytics, marketing, customer support and data enrichment for the purposes and pursuant to the legal bases described above; such service providers comprise companies located in the countries in which we operate [...]
\end{itemize}
\label{quote:salesforce}
\end{myquote}

\begin{myquote}{Reddit privacy policy}
Reddit only shares nonpublic information about you in the following ways. We do not sell this information. [...]

\begin{itemize}
    \item \textit{With our service providers.} We may share information with vendors, consultants, and other service providers who need access to such information to carry out work for us. Their use of personal data will be subject to appropriate confidentiality and security measures. A few examples: (i) payment processors who process transactions on our behalf, (ii) cloud providers who host our data and our services, (iii) third-party ads measurement providers who help us and advertisers measure the performance of ads shown on our Services.
\end{itemize}
\label{quote:reddit}
\end{myquote}

\begin{myquote}{Spotify privacy policy}
We do not sell personal information. We have taken substantial steps to identify and remediate any data sharing arrangements that could constitute a ``sale'' under CCPA.
\label{quote:spotify}
\end{myquote}

\begin{myquote}{LinkedIn privacy policy}
We do not sell personal information, so we don’t have an opt out.
\label{quote:linkedin}
\end{myquote}

\begin{myquote}{CNN privacy policy}
CCPA defines `sale' very broadly. It includes the sharing of California Information in exchange for anything of value. According to this broad definition, in the year before this section was last updated, we may have sold the following categories of California Information to third parties:
\begin{itemize}
    \item Address and other identifiers -- such as name, phone number, postal address, zip code, email address, account name or number, date of birth, driver's license number, payment card numbers, or other similar identifiers
    \item Characteristics of protected classifications -- such as race, ethnicity, or sexual orientation [...]
\end{itemize}
\label{quote:cnn}
\end{myquote}

\begin{myquote}{Hulu privacy policy}
Hulu may disclose certain information about you if you are a registered user of the Hulu services for purposes that may be considered a ``sale'' under CCPA. For example, we may disclose information to advertising partners, advertising technology companies, and companies that perform advertising-related services in order to provide you with more relevant advertising tailored to your interests on the Hulu services.
\label{quote:hulu}
\end{myquote}

\begin{myquote}{Target privacy policy}
\uline{Categories of Personal Information Target has sold about Guests in the preceding 12 months}

\textbf{Categories of Personal Information sold}
\begin{itemize}
    \item Internet or other electronic network activity (e.g. “cookies” or other tracking tags)
\end{itemize}

\textbf{Categories of Third Parties to Whom Personal Information was Sold}
\begin{itemize}
    \item Advertising Networks
\end{itemize}

\uline{Target does not knowingly sell the Personal \\ Information of Minors Under the age of 16.}
\label{quote:target}
\end{myquote}

\begin{myquote}{Healthline privacy policy}
\textbf{Objection to Certain Processing}. When we process information on our own behalf, you may object to our use of your Personal Information by contacting us at the address described below. For example, California residents may be entitled to object, or opt-out, of having their information sold to third parties. If you are a California resident and would like to opt-out of having your information sold, then please visit the following link: \textbf{Do Not Sell My Personal Information}. We do not discriminate against California consumers who exercise any of their rights described in this Policy.
\label{quote:healthline}
\end{myquote}

\begin{myquote}{Twitch privacy policy}
In the twelve months prior to the effective date of this California Notice, Twitch has not sold any personal information of consumers, as those terms are defined under the California Consumer Privacy Act.
\label{quote:twitch}
\end{myquote}

\begin{myquote}{Quizlet privacy policy}
\textbf{Advertising, Analytics and Marketing}
In order to support the Quizlet business and provide our users with relevant information regarding the Service and opportunities presented by partners, Quizlet may use your personal information for marketing the Service. We may also communicate with you and others about opportunities, products, services, contests, promotions, discounts, incentives, surveys, and rewards offered by us and select partners. We may also displaying relevant advertising to you while using the Service. To learn more about how we display relevant advertising to you, and to exercise choices please read the Choices section below and our Ad and Cookie Policy.
\label{quote:quizlet}
\end{myquote}

\begin{myquote}{Medium privacy policy}
Medium does not sell your personal information.

Subject to certain limitations, you have the right to [...] opt out of any sales of your personal information, if we engage in that activity in the future.
\label{quote:medium}
\end{myquote}

\begin{myquote}{Indeed privacy policy}
This list is not meant to be exclusive but shows examples of what our products do. We say ``may'' fall under the definition of ``selling'' data because CCPA is new and these issues are not yet resolved.
\begin{itemize}
    \item We display your public resume to employers and recruiters.
    \item We recommend your public resume to employers and recruiters.
    \item We allow third-party cookies to help employers measure job listing performance and to optimize our online advertising.
    \item We share data with our affiliates as described in our Privacy Center.
\end{itemize}
\label{quote:indeed}
\end{myquote}

\begin{myquote}{eBay privacy policy}
\begin{itemize}
    \item \textbf{Non-registered eBay users (e.g. guest users) or eBay users with suspended accounts}
    \begin{itemize}
        \item For non-registered eBay users and eBay users with suspended accounts, you may submit your deletion request through our Privacy Center webform or email by clicking here. 
        
        \item As outlined on our Privacy Center Contact Page, eBay is required to verify the identity of everyone to whom we release personal information. Please include your first name, last name, country of residence, and the email address you used for any guest transactions you made. Before fulfilling your request, we’ll ask you for POI and POA to ensure that we are only deleting personal information for authorized parties.
        
        \item The POI/POA you provide will be used solely for the purpose of identity verification related to this transaction (i.e. the deletion of personal information) and will be processed and stored in accordance with Section 7 of our User Privacy Notice.
    \end{itemize}
\end{itemize}
\label{quote:ebay}
\end{myquote}

\begin{myquote}{Pinterest privacy policy}
We are also required to communicate information about rights California residents have under California law. You may exercise the following rights: [...]

\begin{itemize}
    \item Right to Delete. You may submit a verifiable request to close your account and we will delete Personal Information about you that we have collected.
\end{itemize}

In order to verify your identity when you make a request, you will be required to log in to your password-protected account or respond to an email verification request.
\label{quote:pinterest}
\end{myquote}

\begin{myquote}{Netflix privacy policy}
If you do not own a Netflix account, we may not be able to respond to requests to exercise rights under CCPA, including the right to know or delete CCPA personal information. Because we only collect limited information about individuals without an account, we are unable to verify requests from non-accountholders to the standard required by CCPA.
\label{quote:netflix}
\end{myquote}

\begin{myquote}{Washington Post privacy policy}
CCPA Rights. California residents can make certain requests about their personal information under the CCPA. Specifically, if you are a California resident, you may request that we: [...]

\begin{itemize}
    \item delete certain information we have about you; and/or provide you with information about the financial incentives that we offer to you, if any.
\end{itemize}

[...] Please note that certain information may be exempt from such requests under applicable law. For example, we may retain certain information for legal compliance and to secure our Services. We may need certain information in order to provide the Services to you; if you ask us to delete it, you may no longer be able to use the Services.

[...] To protect your privacy and security, we take reasonable steps to verify your identity and requests before granting such requests, including by verifying your account information, residency or the email address you provide.
\label{quote:wp}
\end{myquote}

\begin{myquote}{Capital One privacy policy}
California residents also have the right to submit a request for deletion of personal information under certain circumstances, although there may be legal or other reasons that Capital One will retain your information.
\label{quote:co}
\end{myquote}

\begin{myquote}{Walmart privacy policy}
2. Delete My Personal Information: You have the right to ask that we delete your personal information. Once we receive a request, we will delete the personal information (to the extent required by law) we hold about you as of the date of your request from our records and direct any service providers to do the same.
\label{quote:walmart}
\end{myquote}

\begin{myquote}{Best Buy privacy policy}
\textbf{You'll lose your personalized experience with Best Buy:}
The content and offers you see from Best Buy may not be as relevant to you. Also:
\begin{itemize}
    \item You'll no longer have a Best Buy account. If you have an online account, your username and password will be deleted unless you have an active service plan or contract.
    \item You'll lose your order and purchase history.
    \item It's permanent. You can't undo it.
\end{itemize}
\label{quote:bb}
\end{myquote}

\begin{myquote}{Zillow privacy policy}
We provide these tools for your benefit and we will never discriminate against you for using them. But if you choose to delete your data or close your account, we won't be able to offer you services that require us to use your data.
\label{quote:zillow}
\end{myquote}

\section{Internet Archive snapshots for privacy policies}
\label{sec:appendix:archive}

The URLs listed below directly link to Internet Archive snapshots of the pages on each website with CCPA disclosures, as they appeared on the analysis date of February 15, 2021 (i.e., we did not find any evidence that their text had been updated in intervening time). The order is based on the Alexa rankings \cite{Alexa2021} on the same day.

\begin{enumerate}[{[}1{]},itemsep=2pt]
\small
	\item \textbf{Google.com} \url{http://web.archive.org/web/20210215083848/https://policies.google.com/privacy?hl=en-US#enforcement}
	\item \textbf{Youtube.com} See [1]
	\item \textbf{Amazon.com} \url{http://web.archive.org/web/20210228200344/https://www.amazon.com/gp/help/customer/display.html?nodeId=GC5HB5DVMU5Y8CJ2}
	\item \textbf{Yahoo.com} \url{http://web.archive.org/web/20210215085957/https://www.verizonmedia.com/policies/us/en/verizonmedia/privacy/california/index.html}
	\item \textbf{Zoom.us} \url{http://web.archive.org/web/20210122100804/https://zoom.us/docs/en-us/ca-privacy-rights.html}
	\item \textbf{Facebook.com} \url{http://web.archive.org/web/20210213111845/https://www.facebook.com/legal/policy/ccpa}
	\item \textbf{Reddit.com} \url{http://web.archive.org/web/20210215093644/https://www.redditinc.com/policies/privacy-policy}
	\item \textbf{Wikipedia.org} \url{http://web.archive.org/web/20210215102240/https://foundation.wikimedia.org/wiki/Privacy_policy}
	\item \textbf{Myshopify.com} \url{http://web.archive.org/web/20210304194632/https://www.shopify.com/legal/privacy/customers}
	\item \textbf{Ebay.com} \url{http://web.archive.org/web/20210301075255/https://www.ebayinc.com/company/privacy-center/privacy-notice/state-privacy-disclosures/}
	\item \textbf{Office.com} See [15]
	\item \textbf{Live.com} See [15]
	\item \textbf{Netflix.com} \url{http://web.archive.org/web/20210215095542/https://help.netflix.com/legal/privacy#ccpa}
	\item \textbf{Bing.com} See [15]
	\item \textbf{Microsoft.com} \url{http://web.archive.org/web/20210215072740/https://privacy.microsoft.com/en-us/privacystatement#mainotherimportantprivacyinformationmodule}
	\item \textbf{Instructure.com} \url{http://web.archive.org/web/20210217114132/https://www.instructure.com/policies/ccpa}
	\item \textbf{Instagram.com} See [6]
	\item \textbf{Twitch.tv} \url{http://web.archive.org/web/20210301161128/https://www.twitch.tv/p/en/legal/california-privacy-disclosure/}
	\item \textbf{Microsoftonline.com} See [15]
	\item \textbf{Zillow.com} \url{http://web.archive.org/web/20210215055346/https://www.zillowgroup.com/zg-privacy-policy/}
	\item \textbf{Cnn.com} \url{http://web.archive.org/web/20210215083202/https://www.cnn.com/privacy0}
	\stepcounter{enumi}
	\item \textbf{Chase.com} \url{http://web.archive.org/web/20210218040853/https://www.chase.com/digital/resources/privacy-security/privacy/ca-consumer-privacy-act}
	\item \textbf{Intuit.com} \url{http://web.archive.org/web/20210214010007/https://www.intuit.com/privacy/statement/#Country\%20and\%20Region-Specific\%20Terms}
	\item \textbf{Adobe.com} \url{http://web.archive.org/web/20210215175602/https://www.adobe.com/privacy/ca-rights.html}
	\item \textbf{Apple.com} \url{http://web.archive.org/web/20210217055600/https://www.apple.com/legal/privacy/california/}
	\item \textbf{Linkedin.com} \url{https://web.archive.org/web/20210531172117/https://www.linkedin.com/legal/california-privacy-disclosure}
	\item \textbf{Etsy.com} \url{http://web.archive.org/web/20210215080012/https://www.etsy.com/legal/privacy/#rights}
	\item \textbf{Walmart.com} \url{http://web.archive.org/web/20210215043126/https://corporate.walmart.com/privacy-security/walmart-privacy-policy#what-are-your-california-privacy-rights}
	\item \textbf{Dropbox.com} \url{http://web.archive.org/web/20210213194007/https://www.dropbox.com/privacy}
	\item \textbf{Nytimes.com} \url{http://web.archive.org/web/20210215202953/https://www.nytimes.com/privacy/california-notice}
	\item \textbf{Espn.com} \url{http://web.archive.org/web/20210215223631/https://privacy.thewaltdisneycompany.com/en/current-privacy-policy/your-california-privacy-rights/}
	\item \textbf{Twitter.com} \url{http://web.archive.org/web/20210215205721/https://twitter.com/en/privacy}
	\item \textbf{Force.com} See [38]
	\item \textbf{Okta.com} \url{http://web.archive.org/web/20210211014144/https://www.okta.com/privacy-policy/#your-california-privacy-rights-17}
	\item \textbf{Wellsfargo.com} \url{http://web.archive.org/web/20210531173740/https://www.wellsfargo.com/privacy-security/california-consumer-privacy-notice/}
	\item \textbf{Indeed.com} \url{http://web.archive.org/web/20210217210647/https://www.indeed.com/legal?hl=en&redirect=true}
	\item \textbf{Salesforce.com} \url{http://web.archive.org/web/20210216124534/https://www.salesforce.com/company/privacy/full_privacy/?bc=WA#15-additional-disclosures-for-california-residents}
	\item \textbf{Fidelity.com} \url{http://web.archive.org/web/20210531174411/https://communications.fidelity.com/consumer-privacy/california/supplemental-disclosure/}
	\item \textbf{Craigslist.org} \url{http://web.archive.org/web/20210215173249/https://www.craigslist.org/about/privacy.policy}
	\item \textbf{Aliexpress.com} \url{http://web.archive.org/web/20210214082850/https://service.aliexpress.com/page/knowledge?pageId=37&category=1000022028&knowledge=1060015216&language=en}
	\item \textbf{Hulu.com} \url{http://web.archive.org/web/20210216000832/https://www.hulu.com/ca-privacy-rights}
	\item \textbf{Imdb.com} \url{http://web.archive.org/web/20210504134757/https://help.imdb.com/article/issues/G73WRBWXE25UUTHP}
	\item \textbf{Amazonaws.com} \url{http://web.archive.org/web/20210215073820/https://aws.amazon.com/privacy/#Additional_Information_for_Certain_Jurisdictions}
	\item \textbf{Imgur.com} \url{http://web.archive.org/web/20210215035303/https://imgur.com/ccpa}
	\item \textbf{Usps.com} \url{http://web.archive.org/web/20210216155954/https://about.usps.com/who/legal/privacy-policy/full-privacy-policy.htm}
	\item \textbf{Ca.gov} \url{http://web.archive.org/web/20210215161612/https://www.ca.gov/privacy-policy/}
	\item \textbf{Spotify.com} \url{http://web.archive.org/web/20210214180508/https://www.spotify.com/us/legal/California-privacy-disclosure/}
	\item \textbf{Quizlet.com} \url{http://web.archive.org/web/20210212041457/https://quizlet.com/privacy}
	\item \textbf{Heavy.com} \url{http://web.archive.org/web/20210216010022/https://heavy.com/privacy-policy/}
	\item \textbf{Msn.com} See [15]
	\item \textbf{Homedepot.com} \url{http://web.archive.org/web/20210214222102/https://www.homedepot.com/privacy/Privacy_Security#CaliforniaPrivacyRights}
	\item \textbf{Paypal.com} \url{http://web.archive.org/web/20210214231400/https://www.paypal.com/us/webapps/mpp/ua/privacy-full#consumerPrivacy}
	\item \textbf{Bestbuy.com} \url{http://web.archive.org/web/20210218061101/https://www.bestbuy.com/site/privacy-policy/california-privacy-rights/pcmcat204400050063.c?id=pcmcat204400050063}
	\item \textbf{Foxnews.com} \url{http://web.archive.org/web/20210215031800/https://www.foxnews.com/privacy-policy#ccpa}
	\item \textbf{Weather.com} \url{http://web.archive.org/web/20210215073412/https://weather.com/en-US/twc/privacy-policy#us-ccpa-notice-new}
	\item \textbf{Fandom.com} \url{http://web.archive.org/web/20210215034843/https://www.fandom.com/privacy-policy#Additional_provisions_of_this_privacy_policy_that_are_applicable_to_California_residents}
	\item \textbf{Tmall.com} \url{https://web.archive.org/web/20210327120341/https://terms.alicdn.com/legal-agreement/terms/suit_bu1_tmall/suit_bu1_tmall201801031144_60809.html} [Note: we used a translation of the text.]
	\item \textbf{Duckduckgo.com} \url{http://web.archive.org/web/20210215032741/https://duckduckgo.com/privacy}
	\item \textbf{Irs.gov} \url{http://web.archive.org/web/20210215022935/https://www.irs.gov/privacy-disclosure/irs-privacy-policy}
	\item \textbf{Stackoverflow.com} See [96]
	\item \textbf{Tiktok.com} \url{http://web.archive.org/web/20210215032341/https://www.tiktok.com/legal/privacy-policy}
	\item \textbf{Target.com} \url{http://web.archive.org/web/20210216102707/https://www.target.com/c/california-residents-privacy-policy/-/N-m2wjt}
	\item \textbf{Washingtonpost.com} \url{http://web.archive.org/web/20210210213322/https://www.washingtonpost.com/privacy-policy/2011/11/18/gIQASIiaiN_story.html?tid=a_inl_manual#CALIFORNIA}
	\item \textbf{Realtor.com} \url{http://web.archive.org/web/20210216180151/https://www.realtor.com/privacy-policy/#california}
	\item \textbf{Medium.com} \url{http://web.archive.org/web/20210218055729/https://policy.medium.com/medium-privacy-policy-f03bf92035c9?gi=sd}
	\item \textbf{Tradingview.com} \url{http://web.archive.org/web/20210210092122/https://www.tradingview.com/privacy-policy/}
	\item \textbf{Github.com} \url{http://web.archive.org/web/20210125112031/https://docs.github.com/en/github/site-policy/githubs-notice-about-the-california-consumer-privacy-act}
	\item \textbf{Capitalone.com} \url{http://web.archive.org/web/20210515133819/https://www.capitalone.com/privacy/ccpa-disclosure}
	\stepcounter{enumi}
	\item \textbf{Wordpress.com} \url{http://web.archive.org/web/20210215210610/https://wordpress.org/about/privacy/}
	\item \textbf{Breitbart.com} \url{http://web.archive.org/web/20210215013622/https://www.breitbart.com/privacy-policy/}
	\item \textbf{Bankofamerica.com} \url{http://web.archive.org/web/20210217225205/https://www.bankofamerica.com/security-center/ccpa-disclosure/}
	\item \textbf{Ups.com} \url{http://web.archive.org/web/20210216105754/https://www.ups.com/us/en/help-center/legal-terms-conditions/privacy-notice.page#contentBlock-19}
	\item \textbf{Canva.com} \url{http://web.archive.org/web/20210403040612/https://www.canva.com/policies/privacy-policy/}
	\item \textbf{Ameritrade.com} \url{http://web.archive.org/web/20210531214008/https://www.tdameritrade.com/privacy-policies.html}
	\item \textbf{Cnbc.com} \url{http://web.archive.org/web/20210215090123/http://www.nbcuni.com/privacy/full-privacy-policy/}
	\item \textbf{Adp.com} \url{http://web.archive.org/web/20210217092606/https://www.adp.com/privacy.aspx}
	\stepcounter{enumi}
	\item \textbf{Qq.com} \url{http://web.archive.org/web/20210531181841/https://www.imqq.com/mobile/privacy/privacy_En.html}
	\stepcounter{enumi}
	\item \textbf{Fedex.com} \url{http://web.archive.org/web/20210217092006/https://www.fedex.com/en-us/trust-center/privacy.html}
	\item \textbf{Patch.com} \url{http://web.archive.org/web/20210215220008/https://patch.com/privacy#annex-2}
	\item \textbf{Xfinity.com} \url{http://web.archive.org/web/20210216150122/https://www.xfinity.com/privacy/policy/staterights#california}
	\item \textbf{Healthline.com} \url{http://web.archive.org/web/20210409145109/https://www.healthline.com/about/privacy-policy}
	\item \textbf{Soundcloud.com} \url{http://web.archive.org/web/20210215153508/https://soundcloud.com/pages/privacy#notice-to-california-users}
	\stepcounter{enumi}
	\item \textbf{Wayfair.com} \url{http://web.archive.org/web/20210302202926/https://terms.wayfair.io/en-US}
	\item \textbf{Sohu.com} \url{http://web.archive.org/web/20210531182509/https://intro.sohu.com/privacy} [Note: we used a translation of the text.]
	\item \textbf{Businessinsider.com} \url{http://web.archive.org/web/20210215025055/https://www.insider-inc.com/privacy-policy}
	\item \textbf{Bbc.com} \url{http://web.archive.org/web/20210531183425/https://www.bbc.co.uk/usingthebbc/cookies/privacy-policy/}
	\item \textbf{Nih.gov} \url{http://web.archive.org/web/20210215192627/https://www.nih.gov/web-policies-notices}
	\item \textbf{Vimeo.com} \url{https://archive.ph/g2zZR#california_users} [Note: We discovered that the Internet Archive does not properly archive this page due to its dynamic nature. This is a snapshot of the page on a different archival site.]
	\item \textbf{Pinterest.com} \url{http://web.archive.org/web/20210215032949/https://policy.pinterest.com/en/privacy-policy#section-california-residents}
	\item \textbf{Alibaba.com} \url{http://web.archive.org/web/20210216141606/http://rule.alibaba.com/rule/detail/2034.htm}
	\item \textbf{Stackexchange.com} \url{http://web.archive.org/web/20210215134420/https://stackoverflow.com/legal/privacy-policy}
	\item \textbf{Investopedia.com} \url{http://web.archive.org/web/20210215084309/https://www.investopedia.com/legal-4768893#california-privacy-notice}
	\item \textbf{Cnet.com} \url{https://archive.is/vIC7t} \newline
	[Note: We discovered that the Internet Archive does not properly archive this page due to its dynamic nature. This is a snapshot of a newer version of the page on a different archival site; we see no differences in the CCPA disclosures.]
	\item \textbf{Yelp.com} \url{http://web.archive.org/web/20210215053945/https://terms.yelp.com/privacy/en_us/20200101_en_us/#California-Residents:-Your-California-Privacy-Rights}
	\item \textbf{Airbnb.com} \url{http://web.archive.org/web/20210121161940/https://www.airbnb.com/help/article/2863/california-and-vermont}
\end{enumerate}

\section{User survey contents}
\label{sec:appendix:survey}
{\small
\parindent0pt
\renewcommand{\labelitemi}{$\bullet$}
\renewcommand{\labelitemii}{$\circ$}
\renewcommand{\labelitemiii}{\tiny$\blacksquare$}
\renewcommand{\labelitemiv}{\tiny$\blacksquare$}

\subsection{Pre-survey questions}
Which of the following privacy-related laws and regulations have you \textbf{heard of}? Your answers will not affect the rest of the survey.

\medskip
\begin{tabularx}{\columnwidth}{X|X|X|X}
    & \textbf{I have heard of it} & \textbf{I have never heard of it} & \textbf{I'm unsure} \\
    \toprule
    California Consumer Privacy Act (CCPA) & & & \\
    \midrule
    California Privacy Rights Act (CPRA) & & & \\
    \midrule
    California Online Privacy Protection Act (CalOPPA) & & & \\
    \midrule
    European Union General Data Protection Regulation (EU GDPR) & & & \\
    \midrule
    Usability of Digital Privacy Disclosures Act (UDPDA) [\textit{Fake option}] & & & \\
    \midrule
    Nevada Privacy of Information Collected on the Internet from Consumers Act (NPICICA) & & & \\
    \bottomrule
\end{tabularx}

\subsection{Main survey} 
[\textit{Respondents saw exactly one of the sets of questions below; for each set, respondents saw the three excerpts in a randomized order, but they saw the two questions corresponding to each excerpt in the same order. For brevity, we show all answer options on one line; in the actual survey, they are shown as a multiple choice list. After each Likert scale response, we also asked the following two free text response questions, which we again drop from the following text for brevity:}]
\begin{itemize}
\renewcommand{\labelitemi}{$\bullet$}
    \item Please paste one or two sentences from the excerpt you just read that most influenced your answer.
    \item Please briefly explain your answer to the previous question in one or two sentences below.
\end{itemize}

\subsubsection{Instructions} \hspace*{\fill} \\
You will be presented with three excerpts of text from a website's privacy policy, and answer a set of questions for each one. Each question will be \textbf{on the same page} as the relevant excerpt of text from the privacy policy. Note that hyperlinks have been removed from the text.

For each question, make your choice about what you think is \textbf{the correct answer}. \textbf{Answer the questions as if you held an account with the website}. Please take the time to think about your responses carefully; remember, \textbf{we will compensate you with an additional \$0.15 for each question that you answer correctly, plus an extra \$1.00 if you answer all of them correctly}.

\medskip
\noindent
\includegraphics[width=\linewidth]{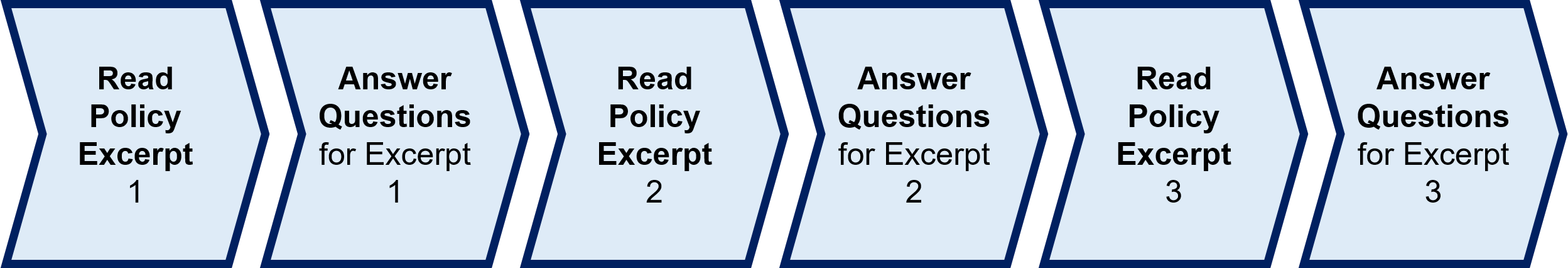}

\subsubsection{Privacy Policy 1: Walmart} 
\hspace*{\fill} \\
\textbf{Please read the following section of text from the privacy policy of Company XYZ, a supermarket, department, and grocery store, and answer the questions below.}
\medskip

\textbf{Categories of Personal Information We Disclose to Each Type of Third Party}

We may share your personal information with certain categories of third parties, as described below.

\medskip
\begin{tabularx}{\columnwidth}{|X|X|}
    \toprule
    \textbf{Types of Third Parties to Which the Personal Information Was Disclosed for a Business Purpose} & \textbf{Categories of Personal Information Disclosed for a Business Purpose} \\
    \midrule
    \multirow[t]{7}{10.5em}{\textbf{Financial service vendors}} & Demographic information \\
    \cline{2-2}
    & Device and online identifiers \\
    \cline{2-2}
    & Financial information \\
    \cline{2-2}
    & Individual preferences and characteristics \\
    \cline{2-2}
    & Location \\
    \cline{2-2}
    & Personal identifiers \\
    \cline{2-2}
    & Purchase history information \\
    \midrule
    \multirow[t]{9}{10em}{\textbf{Service providers that receive data in order to provide services to us (e.g. technology providers, cloud storage providers, etc.)}} & Demographic information \\
    \cline{2-2}
    & Device and online identifiers \\
    \cline{2-2}
    & Employment information \\
    \cline{2-2}
    & Financial information \\
    \cline{2-2}
    & Individual preferences and characteristics \\
    \cline{2-2}
    & Internet, application, and network activity \\
    \cline{2-2}
    & Location information \\
    \cline{2-2}
    & Personal identifiers \\
    \cline{2-2}
    & Purchase history information \\
    \bottomrule
\end{tabularx}
\par

\medskip
\hrulefill
\begin{itemize}
    \item As a California resident, if you make purchases through a website owned by Company XYZ (for example, if you purchase groceries online from Company XYZ and pay with PayPal), is Company XYZ allowed to share your \textbf{past purchase history} with a payment service provider (such as PayPal)?
    \begin{itemize}
        \item Definitely allowed / Probably allowed / I'm really not sure / Probably not allowed / Definitely not allowed
    \end{itemize}
    
    \item As a California resident, if you make purchases through a website owned by Company XYZ (for example, if you purchase groceries online from Company XYZ and pay with PayPal), is Company XYZ allowed to share your \textbf{inferred ethnicity} (inferred based on your purchase history) with a payment service provider (such as PayPal)?
    \begin{itemize}
        \item Definitely allowed / Probably allowed / I'm really not sure / Probably not allowed / Definitely not allowed
    \end{itemize}
    
    \item What was the category of the website in the privacy statement you just read? [\textit{Attention check}]
\end{itemize}
\hrulefill
\medskip

\textbf{Please read the following section of text from the privacy policy of Company XYZ, a supermarket, department, and grocery store, and answer the questions below.}

\textit{Note that hyperlinks have been removed from this excerpt.}
\medskip

\textbf{Stop Selling My Personal Information}: We do not share your personal information for money. We do share personal information within our family of companies and with certain third parties, which may be considered a ``sale'' under California law even if we don't receive money in exchange for the information. For purposes of California law, we may share with these parties all categories of personal information, except for background and criminal information, biometric information, and government identifiers. We do not knowingly sell the personal information of minors under 16 years of age. If you are a California resident, you have a right to opt-out of the sale of your personal information. To do so, click ``Do Not Sell My Personal Information''.

\medskip
We will not discriminate against you for exercising your rights. This generally means we will not deny you goods or services, charge different prices or rates, provide a different level of service or quality of goods, or suggest that you might receive a different price or level of quality for goods. Please know, if you ask us to stop selling your data, it may impact your experience with us, and you may not be able to participate in certain programs or membership services which require usage of your personal information to function.

\hrulefill
\begin{itemize}
    \item As a California resident, if you opt out of the sale of your personal information as described, is Company XYZ allowed to provide your \textbf{full name and mailing address} to an advertising company, \textbf{in return for a portion of revenue} from this advertising company?
    \begin{itemize}
        \item Definitely allowed / Probably allowed / I'm really not sure / Probably not allowed / Definitely not allowed
    \end{itemize}
    
    \item As a California resident, if you opt out of the sale of your personal information as described, is Company XYZ allowed to provide your \textbf{full name and mailing address} to an advertising company, \textbf{without receiving revenue in return} from this advertising company?
    \begin{itemize}
        \item Definitely allowed / Probably allowed / I'm really not sure / Probably not allowed / Definitely not allowed
    \end{itemize}
    
    \item Which of the following categories of personal information did the previous question involve? [\textit{Attention check}]
\end{itemize}
\hrulefill
\medskip

\textbf{Please read the following section of text from the privacy policy of Company XYZ, a supermarket, department, and grocery store, and answer the questions below.}

\textit{Note that hyperlinks have been removed from this excerpt.}
\medskip

\textbf{Delete My Personal Information}: You have the right to ask that we delete your personal information. Once we receive a request, we will delete the personal information (to the extent required by law) we hold about you as of the date of your request from our records and direct any service providers to do the same. In some cases, deletion may be accomplished through de-identification of the information. Choosing to delete your personal information may impact your ability to use our websites and online features, including closure of your online account, and limit your use of in-store functions that require your personal information.

\medskip
We will not discriminate against you for exercising your rights. This generally means we will not deny you goods or services, charge different prices or rates, provide a different level of service or quality of goods, or suggest that you might receive a different price or level of quality for goods. Please know, if you ask us to delete your data, it may impact your experience with us, and you may not be able to participate in certain programs or membership services which require usage of your personal information to function.

\medskip
To exercise the California privacy rights described above, please click ``Request My Personal Information'' or call 1-555-XYZ-mart.

\medskip
\textbf{Verifying Your Identity for Requests to Delete Personal Information}

We take the privacy of your personal information seriously and want to ensure that we provide only you or your authorized agent with your personal information. The California Consumer Privacy Act (CCPA) also requires that we verify the identity of each person who makes a request to delete the personal information we have about you.

\medskip
\textbf{How Do You Verify My Identity?}

We may verify your identity in a few different ways. We've partnered with a third party identification service to help us verify your identity and prevent fraudulent requests. When you make your request, you will be asked to answer a few questions about yourself to help us validate your identity. If you have an online account, we may ask you to log into your account and complete a one-time passcode validation.

\medskip
In some instances, we may ask you to provide other documentation to verify your identity. If this happens, we will reach out to you directly with this request.

\medskip
\textbf{What If You Can't Verify My Identity?}

If we can't verify your identity, we will not be able to process your request to delete the personal information we have about you. If we are unable to verify your identity with a high degree of certainty, we may not be able to delete some of your information.

\hrulefill
\begin{itemize}
    \item As a California resident, if you request the deletion of your personal information as described, is Company XYZ required to direct \textbf{any marketing company that has received your information, either by sharing or selling, from Company XYZ} to delete information about your activity on Company XYZ's websites?
    \begin{itemize}
        \item Definitely required / Probably required / I'm really not sure / Probably not required / Definitely not required
    \end{itemize}
    
    \item As a California resident, if you request the deletion of your personal information as described, is Company XYZ required to direct \textbf{any marketing company that provides advertising consultation services to Company XYZ} (that is, acting in the capacity of a service provider) to delete information about your activity on Company XYZ's websites?
    \begin{itemize}
        \item Definitely required / Probably required / I'm really not sure / Probably not required / Definitely not required
    \end{itemize}
\end{itemize}
\hrulefill

\subsubsection{Privacy Policy 2: eBay} 
\hspace*{\fill} \\
\textbf{Please read the following section of text from the privacy policy of Company XYZ, an online shopping and auction platform, and answer the questions below.}
\medskip

We share personal information with service providers and other vendors. The following Table lists the categories of personal information disclosed for a business purpose and the categories of recipients who have received this personal information. Please note that not every one of the following categories may apply to your personal information.

\medskip
\begin{tabularx}{\columnwidth}{|X|X|}
    \toprule
    \textbf{Category of Personal Information Disclosed for a Business Purpose} & \textbf{Categories of Recipients} \\
    \midrule
    \begin{enumerate}[leftmargin=0.5cm,topsep=0pt]
        \item Identifiers
        \item Customer Records
        \item Personal Characteristics or Traits
        \item Customer Account Details / Commercial Information
        \item Internet Usage Information
        \item Geolocation Data
        \item Inferences derived from Personal Information
    \end{enumerate} &
    \begin{itemize}[leftmargin=0.5cm,topsep=0pt]
        \item Service providers
        \item Credit agencies (when required by law) 
        \item Participants in the Company XYZ xRewards program 
        \item Financial institutions
        \item Government authorities and agencies
    \end{itemize} \\
    \bottomrule
\end{tabularx}
\par
\medskip

\hrulefill
\begin{itemize}
    \item As a California resident, if you make purchases through a website owned by Company XYZ (for example, if you win a bid on an auction listing hosted by Company XYZ and pay with PayPal), is Company XYZ allowed to share your \textbf{past purchase history} with a payment service provider (such as PayPal)?
    \begin{itemize}
        \item Definitely allowed / Probably allowed / I'm really not sure / Probably not allowed / Definitely not allowed
    \end{itemize}
    
    \item As a California resident, if you make purchases through a website owned by Company XYZ (for example, if you win a bid on an auction listing hosted by Company XYZ and pay with PayPal), is Company XYZ allowed to share your \textbf{inferred ethnicity} (inferred based on your purchase history) with a payment service provider (such as PayPal)?
    \begin{itemize}
        \item Definitely allowed / Probably allowed / I'm really not sure / Probably not allowed / Definitely not allowed
    \end{itemize}
    
    \item What was the category of the website in the privacy statement you just read? [\textit{Attention check}]
\end{itemize}
\hrulefill
\medskip

\textbf{Please read the following section of text from the privacy policy of Company XYZ, an online shopping and auction platform, and answer the questions below.}

\textit{Note that hyperlinks have been removed from this excerpt.}

\begin{itemize}
    \item \textbf{The right to opt-out}
    \begin{itemize}
        \item California consumers have the right to opt-out of any ``sale'' of their personal information. As mentioned, the California Consumer Privacy Act (CCPA)'s definition of ``sale'' is broad and includes sharing practices common among online marketplaces. While, without your consent, Company XYZ does not sell personal information for money, we do share information for a variety of other purposes, such as to tailor your online experience or provide more relevant advertisements to you.
    \end{itemize}
    
    \item \textbf{How to exercise the right to opt-out}
    \begin{itemize}
        \item \textbf{Registered Company XYZ users with accounts in \\ good standing}
        \begin{itemize}
            \item Click here to go to our opt-out page where you can opt-out.
            \item Alternatively, you may submit your request through our Privacy Center webform or email by clicking here.
            \begin{itemize}
                \item Under “How can we help you?”, select “Other privacy request.”
                \item In the comment box, type “CCPA opt-out request.”
                \item In addition, you may contact Customer Service to submit your opt-out request.
            \end{itemize}
        \end{itemize}
    \end{itemize}
\end{itemize}

\hrulefill
\newpage
\begin{itemize}
    \item As a California resident, if you opt out of the sale of your personal information as described, is Company XYZ allowed to provide your \textbf{full name and mailing address} to an advertising company, \textbf{in return for a portion of revenue} from this advertising company?
    \begin{itemize}
        \item Definitely allowed / Probably allowed / I'm really not sure / Probably not allowed / Definitely not allowed
    \end{itemize}
    
    \item As a California resident, if you opt out of the sale of your personal information as described, is Company XYZ allowed to provide your \textbf{full name and mailing address} to an advertising company, \textbf{without receiving revenue in return} from this advertising company?
    \begin{itemize}
        \item Definitely allowed / Probably allowed / I'm really not sure / Probably not allowed / Definitely not allowed
    \end{itemize}
    
    \item Which of the following categories of personal information did the previous question involve? [\textit{Attention check}]
\end{itemize}
\hrulefill
\medskip

\textbf{Please read the following section of text from the privacy policy of Company XYZ, an online shopping and auction platform, and answer the questions below.}

\textit{Note that hyperlinks have been removed from this excerpt.}

\begin{itemize}
    \item \textbf{The right to deletion}
    \begin{itemize}
        \item Subject to some exceptions, California consumers have the right to request the deletion of the personal information we have collected from them.
        
        \item Deletion requests may take between 30 and 60 days to fulfill (depending on whether you have recently sold on Company XYZ as we must meet our commitment under Company XYZ's Money Back Guarantee policy). If we need more time or additional information to fulfill your request, we will write to let you know.
    \end{itemize}
    
    \item \textbf{How to exercise the right to opt-out}
    \begin{itemize}
        \item \textbf{Registered Company XYZ users with accounts in \\ good standing}
        \begin{itemize}
            \item If you are a registered Company XYZ user with an account in good standing, you can submit a deletion request at any time through your Company XYZ account. Your account will be closed as part of fulfilling your deletion request. We will verify you using your existing username and password prior to allowing you to make a request.

            \item Click here to submit your request. You can also learn more about deleting your data by visiting our Privacy Center Contact Page.
        \end{itemize}
    \end{itemize}
\end{itemize}

\hrulefill
\begin{itemize}
    \item As a California resident, if you request the deletion of your personal information as described, is Company XYZ required to direct \textbf{any marketing company that has received your information, either by sharing or selling, from Company XYZ} to delete information about your activity on Company XYZ's websites?
    \begin{itemize}
        \item Definitely required / Probably required / I'm really not sure / Probably not required / Definitely not required
    \end{itemize}
    
    \item As a California resident, if you request the deletion of your personal information as described, is Company XYZ required to direct \textbf{any marketing company that provides advertising consultation services to Company XYZ} (that is, acting in the capacity of a service provider) to delete information about your activity on Company XYZ's websites?
    \begin{itemize}
        \item Definitely required / Probably required / I'm really not sure / Probably not required / Definitely not required
    \end{itemize}
\end{itemize}
\hrulefill

\subsubsection{Privacy Policy 3: The New York Times}
\hspace*{\fill} \\
\textbf{Please read the following section of text from the privacy policy of Company XYZ, a publisher of newspapers and digital news, and answer the questions below.}

\medskip
\textbf{Categories of Personal Information We Collect}
\begin{enumerate}
    \item Identifiers – including name, address, email address, account name, IP address, cookie ID, mobile advertising ID and an ID number assigned to your account.

    \item Other personal information – including phone number, billing address, credit or debit card information, employment or education information.
    
    This category includes personal information protected under pre-existing California law (Cal. Civ. Code 1798.80(e)) and overlaps with other categories.

    \item Demographic information – including your age or gender.
    
    This type of personal information includes what is also considered a protected classification characteristic under pre-existing California or federal laws.

    \item Commercial information on your interactions with the XYZ News Services – including purchases and other commercial engagements with Company XYZ.

    \item Internet or other electronic network activity information – including browsing activity on our sites and apps, browser type and browser language. 

    This also encompasses other information that gets collected automatically when you use our sites and apps or interact with us through social media.

    \item Geolocation data, inferred from your IP address, to help us deliver relevant content and enhance your experience.

    \item Inferences drawn from any of the above personal information – about your preferences, predispositions and behavior as they relate to our sites and apps.
\end{enumerate}

\textbf{The Purposes for Our Collection}

\medskip
We collect the above information for the business purpose of giving you access to our sites and apps. This information is a key part of how we carry out subscription services, account access, customer service, orders and transactions, customer research and feedback programs, analytics, advertising and marketing services.

\medskip
For commercial purposes, we collect information from the following categories: identifiers, demographic information, commercial information, internet activity, geolocation data (based on your IP address but not your precise GPS location).

\hrulefill
\begin{itemize}
    \item As a California resident, if you make purchases through a website owned by Company XYZ (for example, if you purchase a newspaper subscription and pay with PayPal), is Company XYZ allowed to share your \textbf{past purchase history} with a payment service provider (such as PayPal)?
    \begin{itemize}
        \item Definitely allowed / Probably allowed / I'm really not sure / Probably not allowed / Definitely not allowed
    \end{itemize}
    
    \item As a California resident, if you make purchases through a website owned by Company XYZ (for example, if you purchase a newspaper subscription and pay with PayPal), is Company XYZ allowed to share your \textbf{inferred ethnicity} (inferred based on your purchase history) with a payment service provider (such as PayPal)?
    \begin{itemize}
        \item Definitely allowed / Probably allowed / I'm really not sure / Probably not allowed / Definitely not allowed
    \end{itemize}
    
    \item What was the category of the website in the privacy statement you just read? [\textit{Attention check}]
\end{itemize}
\hrulefill
\medskip

\textbf{Please read the following section of text from the privacy policy of Company XYZ, a publisher of newspapers and digital news, and answer the questions below.}

\medskip
Company XYZ does not sell personal information of its readers as the term ``sell'' is traditionally understood. But ``sell'' under the California Consumer Privacy Act (CCPA) is broadly defined. It includes the sharing of personal information with third parties in exchange for something of value, even if no money changes hands. For example, sharing an advertising or device identifier to a third party may be considered a ``sale'' under the CCPA.

\medskip
To the extent Company XYZ ``sells'' your personal information (as the term ``sell'' is defined under the CCPA), you have the right to opt-out of that ``sale'' on a going-forward basis at any time.

\medskip
Once you have opted out, you will see a change to "We No Longer Sell Your Personal Information." If you have an account with certain XYZ News Services and are logged in, we will save your preference and honor your opt-out request across browsers and devices so long as you remain logged in. If you are not logged in, or do not have an account with any XYZ News Services listed above, your opt-out of the ``sale'' of personal information will be specific to the browser or device from which you have clicked ``Do Not Sell My Personal Information'' and until you clear your cookies (or local storage in apps) on this browser or device.

\medskip
After you opt out of the ``sale'' of your personal information, we will no longer ``sell'' your personal information to third parties (except in an aggregated or de-identified manner so it is no longer personal information), but we will continue to share your personal information with our service providers, which process it on our behalf. Exercising your right to opt out of the ``sale'' of your personal information does not mean that you will stop seeing ads on our sites and apps.

\hrulefill
\begin{itemize}
    \item As a California resident, if you opt out of the sale of your personal information as described, is Company XYZ allowed to provide your \textbf{full name and mailing address} to an advertising company, \textbf{in return for a portion of revenue} from this advertising company?
    \begin{itemize}
        \item Definitely allowed / Probably allowed / I'm really not sure / Probably not allowed / Definitely not allowed
    \end{itemize}
    
    \item As a California resident, if you opt out of the sale of your personal information as described, is Company XYZ allowed to provide your \textbf{full name and mailing address} to an advertising company, \textbf{without receiving revenue in return} from this advertising company?
    \begin{itemize}
        \item Definitely allowed / Probably allowed / I'm really not sure / Probably not allowed / Definitely not allowed
    \end{itemize}
    
    \item Which of the following categories of personal information did the previous question involve? [\textit{Attention check}]
\end{itemize}
\hrulefill
\medskip

\textbf{Please read the following section of text from the privacy policy of Company XYZ, a publisher of newspapers and digital news, and answer the questions below.}

\textit{Note that hyperlinks have been removed from this excerpt.}

\medskip
In some parts of the world, you have the right to access, modify, or delete the personal information we have about you.

\medskip
If you'd like to exercise any of the above rights, contact us via this form or by calling us at our toll-free number, 1-555-616-XYZN. In your request, please be specific. State the information you want changed, whether you'd like your information suppressed from our database or whether there are limitations you'd like us to put on how we use your personal information. Please use the email address linked to that personal information — we only complete requests on the information linked to your email address. To verify your identity, we will email the email address you provide us, and which matches our records, and wait for your response. In some instances we may also ask for additional information. This is how we verify your identity before complying.

\medskip
We might need to keep certain information for recordkeeping purposes, or to complete a transaction you began prior to requesting a change or deletion (e.g., if you make a purchase or enter a promotion, you might not be able to change or delete the personal information provided until after the completion of the purchase or promotion).

\medskip
In some cases, your request doesn't ensure complete removal of the content or information (e.g., if another user has reposted your content).

\hrulefill
\begin{itemize}
    \item As a California resident, if you request the deletion of your personal information as described, is Company XYZ required to direct \textbf{any marketing company that has received your information, either by sharing or selling, from Company XYZ} to delete information about your activity on Company XYZ's websites?
    \begin{itemize}
        \item Definitely required / Probably required / I'm really not sure / Probably not required / Definitely not required
    \end{itemize}
    
    \item As a California resident, if you request the deletion of your personal information as described, is Company XYZ required to direct \textbf{any marketing company that provides advertising consultation services to Company XYZ} (that is, acting in the capacity of a service provider) to delete information about your activity on Company XYZ's websites?
    \begin{itemize}
        \item Definitely required / Probably required / I'm really not sure / Probably not required / Definitely not required
    \end{itemize}
\end{itemize}
\hrulefill

\subsubsection{Privacy Policy 4: Hulu} 
\hspace*{\fill} \\
\textbf{Please read the following section of text from the privacy policy of Company XYZ, a streaming service, and answer the questions below.}

\medskip
We collect the following categories of personal information, as defined in the California Consumer Privacy Act of 2018 (``CCPA''), from or about you and may have disclosed this information for a business purpose:
\renewcommand{\labelitemi}{\tiny$\blacksquare$}
\begin{itemize}
    \item Identifiers, such as your name, IP address, email, and other similar identifiers;

    \item Personal information categories listed in the California Customer Records provision, such as payment information;

    \item Characteristics about you, such as age and gender;

    \item Commercial information, such as the XYZ Streaming Service plan(s) you purchase;

    \item Internet or other electronic network activity information, such as your viewing activity on the XYZ Streaming Service;

    \item Inferences based on the information we collect from or about you which we use to better understand your preferences and behavior, including to customize Content and advertising for you.
\end{itemize}
\renewcommand{\labelitemi}{$\bullet$}

We may share information collected from or about you with others, including business partners, social networking services, service providers, advertisers, and other companies that are not affiliated with Company XYZ, for the purposes described below.

\medskip
\uline{Service Providers}. We may share information collected from or about you with companies that provide services to us, our business partners, and advertisers, including companies that assist with payment processing, analytics, data processing and management (e.g., to facilitate our targeted advertising and marketing efforts), account management, hosting, customer and technical support, marketing, advertising, measurement, and other services. The categories of information that we share with service providers will vary depending on the types of services they provide. For example, we share payment information with our third-party payment processor to assist us in our billing efforts, email addresses with our marketing service providers to assist with our email campaigns, and IP address, hashed emails, and advertising identifiers with service providers to assist in our advertising and marketing efforts.

\hrulefill
\begin{itemize}
    \item As a California resident, if you make purchases through a website owned by Company XYZ (for example, if you make a monthly payment for your XYZ Streaming Service subscription with PayPal), is Company XYZ allowed to share your \textbf{past purchase history} with a payment service provider (such as PayPal)?
    \begin{itemize}
        \item Definitely allowed / Probably allowed / I'm really not sure / Probably not allowed / Definitely not allowed
    \end{itemize}
    
    \item As a California resident, if you make purchases through a website owned by Company XYZ (for example, if you make a monthly payment for your XYZ Streaming Service subscription with PayPal), is Company XYZ allowed to share your \textbf{inferred ethnicity} (inferred based on your purchase history) with a payment service provider (such as PayPal)?
    \begin{itemize}
        \item Definitely allowed / Probably allowed / I'm really not sure / Probably not allowed / Definitely not allowed
    \end{itemize}
    
    \item What was the category of the website in the privacy statement you just read? [\textit{Attention check}]
\end{itemize}
\hrulefill
\medskip

\textbf{Please read the following section of text from the privacy policy of Company XYZ, a streaming service, and answer the questions below.}

\textit{Note that hyperlinks have been removed from this excerpt.}

\medskip
Beginning January 1, 2020, the California Consumer Privacy Act of 2018 (``CCPA'') provides California residents with additional rights as described below.

\renewcommand{\labelitemi}{$\circ$}
\renewcommand{\labelitemii}{\tiny$\blacksquare$}
\begin{itemize}
    \item \textit{Right to Opt Out of ``Sale''.} Company XYZ may disclose certain information about you if you are a registered user of the XYZ Streaming Service services for purposes that may be considered a ``sale'' under the CCPA. For example, we may disclose information to advertising partners, advertising technology companies, and companies that perform advertising-related services in order to provide you with more relevant advertising tailored to your interests on the Company XYZ services. This information may include identifiers such as your IP address, advertising identifiers, your email address (in a de-identified or hashed form), age and gender, your internet or other electronic network information such as your interaction with an ad, and geolocation data. We may also disclose to our content programmers information about you, which may help personalize your experience and the content and ads you see on the XYZ Streaming Service as well as other platforms and services, as further described in our Privacy Policy. To learn more, including about your opt-out choices, please click here.

    \medskip
    \item \textit{How to Exercise Your CCPA Rights}
    \begin{itemize}
        \item We provide these rights, subject to a verifiable consumer request, to account holders regarding personal information collected by us in connection with your account, including any profiles. To submit a request to exercise these rights, log into your account on the XYZ Streaming Service website, go to ``Privacy and Settings'' in account settings and click on ``California Privacy Rights'' and follow the instructions. You will be required to verify your request via your XYZ Streaming Service account credentials and/or by other means. You can also call our toll-free number at 555-873-2637 to speak with one of our representatives.

        \item If you are not a registered user of the XYZ Streaming Service services and do not have XYZ Streaming Service account credentials, please note we have limited information about you. You will be required to verify your request, including by providing your name and the email that you believe we have collected about you. Please click here to make a request.
    \end{itemize}
\end{itemize}
\renewcommand{\labelitemi}{$\bullet$}
\renewcommand{\labelitemii}{$\circ$}

\hrulefill
\begin{itemize}
    \item As a California resident, if you opt out of the sale of your personal information as described, is Company XYZ allowed to provide \textbf{your full name and mailing address} to \textbf{an advertising company} so that the advertising company can display advertisements to you on other websites, \textbf{without receiving revenue in return} from this advertising company?
    \begin{itemize}
        \item Definitely allowed / Probably allowed / I'm really not sure / Probably not allowed / Definitely not allowed
    \end{itemize}
    
    \item As a California resident, if you opt out of the sale of your personal information as described, is Company XYZ allowed to provide your \textbf{full name and mailing address} to \textbf{a separately-owned streaming service} so that the two companies can provide joint product offers to you, \textbf{without receiving revenue in return} from this partner company?
    \begin{itemize}
        \item Definitely allowed / Probably allowed / I'm really not sure / Probably not allowed / Definitely not allowed
    \end{itemize}
    
    \item Which of the following categories of personal information did the previous question involve? [\textit{Attention check}]
\end{itemize}
\hrulefill
\medskip

\textbf{Please read the following section of text from the privacy policy of Company XYZ, a streaming service, and answer the questions below.}

\textit{Note that hyperlinks have been removed from this excerpt.}

\medskip
Beginning January 1, 2020, the California Consumer Privacy Act of 2018 (``CCPA'') provides California residents with additional rights as described below. Please note your right to delete is subject to certain exceptions under the CCPA.

\renewcommand{\labelitemi}{$\circ$}
\renewcommand{\labelitemii}{\tiny$\blacksquare$}
\begin{itemize}
    \item \textit{Right to Delete.} You have the right to request the deletion of personal information that we collect or maintain about you and your account. Please note that requesting deletion will require the cancellation and deletion of your account which, upon completion, cannot be undone.

    \medskip
    \item \textit{How to Exercise Your CCPA Rights}
    \begin{itemize}
        \item We provide these rights, subject to a verifiable consumer request, to account holders regarding personal information collected by us in connection with your account, including any profiles. To submit a request to exercise these rights, log into your account on the XYZ Streaming Service website, go to ``Privacy and Settings'' in account settings and click on ``California Privacy Rights'' and follow the instructions. You will be required to verify your request via your XYZ Streaming Service account credentials and/or by other means. You can also call our toll-free number at 555-873-2637 to speak with one of our representatives.

        \item If you are not a registered user of the XYZ Streaming Service services and do not have XYZ Streaming Service account credentials, please note we have limited information about you. You will be required to verify your request, including by providing your name and the email that you believe we have collected about you. Please click here to make a request.
    \end{itemize}
\end{itemize}
\renewcommand{\labelitemi}{$\bullet$}
\renewcommand{\labelitemii}{$\circ$}

\hrulefill
\begin{itemize}
    \item As a California resident, if you request the deletion of your personal information as described, is Company XYZ required to direct \textbf{any marketing company that has received your information, either by sharing or selling, from Company XYZ} to delete information about your activity on Company XYZ's websites?
    \begin{itemize}
        \item Definitely required / Probably required / I'm really not sure / Probably not required / Definitely not required
    \end{itemize}
    
    \item As a California resident, if you request the deletion of your personal information as described, is Company XYZ required to direct \textbf{any marketing company that provides advertising consultation services to Company XYZ} (that is, acting in the capacity of a service provider) to delete information about your activity on Company XYZ's websites?
    \begin{itemize}
        \item Definitely required / Probably required / I'm really not sure / Probably not required / Definitely not required
    \end{itemize}
\end{itemize}
\hrulefill

\subsubsection{Privacy Policy 5: ESPN} 
\hspace*{\fill} \\
\textbf{Please read the following section of text from the privacy policy of Company XYZ, a sports news website that operates a streaming service, and answer the questions below.}

\textit{Note that hyperlinks have been removed from this excerpt.}

\medskip
We collected the following categories of personal information in the last 12 months: identifiers/contact information, demographic information (such as gender and age), payment card information associated with you, commercial information, Internet or other electronic network activity information, geolocation data, and inferences drawn from the above.

\medskip
We disclosed the following categories of personal information for a business purpose in the last 12 months: identifiers/contact information, demographic information (such as gender and age), payment card information associated with you, commercial information, Internet or other electronic network activity information, geolocation data, and inferences drawn from the above. We disclosed each category to third-party business partners and service providers, third-party sites or platforms such as social networking sites, and other third parties as described in the ``Sharing Your Personal Information with Other Entities'' section of the Privacy Policy.

\hrulefill
\begin{itemize}
    \item As a California resident, if you make purchases through a website owned by Company XYZ (for example, if you purchase an annual subscription to the XYZ Streaming Service and pay with PayPal), is Company XYZ allowed to share your \textbf{past purchase history} with a payment service provider (such as PayPal)?
    \begin{itemize}
        \item Definitely allowed / Probably allowed / I'm really not sure / Probably not allowed / Definitely not allowed
    \end{itemize}
    
    \item As a California resident, if you make purchases through a website owned by Company XYZ (for example, if you purchase an annual subscription to the XYZ Streaming Service and pay with PayPal), is Company XYZ allowed to share your \textbf{inferred ethnicity} (inferred based on your purchase history) with a payment service provider (such as PayPal)?
    \begin{itemize}
        \item Definitely allowed / Probably allowed / I'm really not sure / Probably not allowed / Definitely not allowed
    \end{itemize}
    
    \item What was the category of the website in the privacy statement you just read? [\textit{Attention check}]
\end{itemize}
\hrulefill
\medskip

\textbf{Please read the following section of text from the privacy policy of Company XYZ, a sports news website that operates a streaming service, and answer the questions below.}

\textit{Note that hyperlinks have been removed from this excerpt.}

\medskip
As the term is defined by the California Consumer Privacy Act (CCPA), we ``sold'' the following categories of personal information in the last 12 months: identifiers/contact information, Internet or other electronic network activity information, and inferences drawn from the above. We ``sold'' each category to advertising networks, data analytics providers, and social networks.

\medskip
The business or commercial purposes of ``selling'' personal information is for third-party companies to perform services on our behalf, such as marketing, advertising, and audience measurement. We do not ``sell'' personal information of known minors under 16 years of age.

\medskip
\textbf{Right to Opt Out of Sale of Personal Information}

If you are a California resident, you have the right to ``opt out'' of the ``sale'' of your ``personal information'' to ``third parties'' (as those terms are defined in the CCPA).

\medskip
\textbf{Process to Make a CCPA Request}

Making Requests to ``Opt Out'' of the ``Sale'' of ``Personal Information''

To submit a request to opt out of the sale of your personal information, you may visit our ``Do Not Sell My Personal Information'' Rights page or send an email to caprivacy@xyzsports.com with the subject line "do not sell info." You may also use an authorized agent to submit a request to opt out on your behalf if you provide the authorized agent signed written permission to do so. Authorized agents may submit requests to opt out by sending an email to caprivacy@xyzsports.com with the subject line "do not sell info."

\medskip
You have the right not to receive discriminatory treatment for the exercise of your privacy rights conferred by the CCPA.

\hrulefill
\begin{itemize}
    \item As a California resident, if you opt out of the sale of your personal information as described, is Company XYZ allowed to provide \textbf{your full name and mailing address} to \textbf{an advertising company} so that the advertising company can display advertisements to you on other websites, \textbf{without receiving revenue in return} from this advertising company?
    \begin{itemize}
        \item Definitely allowed / Probably allowed / I'm really not sure / Probably not allowed / Definitely not allowed
    \end{itemize}
    
    \item As a California resident, if you opt out of the sale of your personal information as described, is Company XYZ allowed to provide your \textbf{full name and mailing address} to \textbf{a separately-owned news website} so that the two companies can provide joint product offers to you, \textbf{without receiving revenue in return} from this partner company?
    \begin{itemize}
        \item Definitely allowed / Probably allowed / I'm really not sure / Probably not allowed / Definitely not allowed
    \end{itemize}
    
    \item Which of the following categories of personal information did the previous question involve? [\textit{Attention check}]
\end{itemize}
\hrulefill
\medskip

\textbf{Please read the following section of text from the privacy policy of Company XYZ, a sports news website that operates a streaming service, and answer the questions below.}

\textit{Note that hyperlinks have been removed from this excerpt.}

\medskip
\textbf{Right to Delete}

If you are a California resident, you have the right to request that we delete personal information that we collect from you, subject to applicable legal exceptions.

\medskip
\textbf{Process to Make a California Consumer Privacy Act (CCPA) Request}

Making Deletion Requests

To make a deletion request, please visit ccpa.xyzsports.com. Before completing your request, we may need to verify your identity. We will send you a link to verify your email address and may request additional documentation or information solely for the purpose of verifying your identity.

\medskip
You have the right not to receive discriminatory treatment for the exercise of your privacy rights conferred by the CCPA.

\hrulefill
\begin{itemize}
    \item As a California resident, if you request the deletion of your personal information as described, is Company XYZ required to direct \textbf{any marketing company that has received your information, either by sharing or selling, from Company XYZ} to delete information about your activity on Company XYZ's websites?
    \begin{itemize}
        \item Definitely required / Probably required / I'm really not sure / Probably not required / Definitely not required
    \end{itemize}
    
    \item As a California resident, if you request the deletion of your personal information as described, is Company XYZ required to direct \textbf{any marketing company that provides advertising consultation services to Company XYZ} (that is, acting in the capacity of a service provider) to delete information about your activity on Company XYZ's websites?
    \begin{itemize}
        \item Definitely required / Probably required / I'm really not sure / Probably not required / Definitely not required
    \end{itemize}
\end{itemize}
\hrulefill

\subsubsection{Privacy Policy 6: Best Buy} 
\hspace*{\fill} \\
\textbf{Please read the following section of text from the privacy policy of Company XYZ, an electronics retailer, and answer the questions below.}

\medskip
During the 12-month period prior to the effective date of this Addendum, we may have:
\begin{enumerate}[a.]
    \setcounter{enumi}{2}
    \item Shared your personal information with the following categories of third parties:
    \begin{itemize}
        \item Our affiliates
        \item Our joint marketing partners
        \item Our business partners
        \item Social media networks
        \item Third-party marketing partners
        \item Government entities, including law enforcement
    \end{itemize}

    \setcounter{enumi}{4}
    \item Disclosed for a business purpose the following categories of personal information about you:
    \begin{itemize}
        \item Identifiers
        \item Identifiers (Online)
        \item Information Related to Characteristics Protected Under California or Federal Law
        \item Commercial Information
        \item Internet and Other Electronic Network Activity Information
        \item Geolocation Data
        \item Profile Inferences
    \end{itemize}
\end{enumerate}

\hrulefill
\begin{itemize}
    \item As a California resident, if you make purchases through a website owned by Company XYZ (for example, if you purchase a mobile phone from Company XYZ and pay with PayPal), is Company XYZ allowed to share your \textbf{past purchase history} with a payment service provider (such as PayPal)?
    \begin{itemize}
        \item Definitely allowed / Probably allowed / I'm really not sure / Probably not allowed / Definitely not allowed
    \end{itemize}
    
    \item As a California resident, if you make purchases through a website owned by Company XYZ (for example, if you purchase a mobile phone from Company XYZ and pay with PayPal), is Company XYZ allowed to share your \textbf{inferred ethnicity} (inferred based on your purchase history) with a payment service provider (such as PayPal)?
    \begin{itemize}
        \item Definitely allowed / Probably allowed / I'm really not sure / Probably not allowed / Definitely not allowed
    \end{itemize}
    
    \item What was the category of the website in the privacy statement you just read? [\textit{Attention check}]
\end{itemize}
\hrulefill
\medskip

\textbf{Please read the following section of text from the privacy policy of Company XYZ, an electronics retailer, and answer the questions below.}

\textit{Note that hyperlinks have been removed from this excerpt.}

\medskip
\textbf{Company XYZ does not sell (as ``sell'' is traditionally defined) your personal information.}

That is, we don't provide your name, phone number, address, email address or other personally identifiable information to third parties in exchange for money.

\medskip
But under California law, sharing information for advertising purposes may be considered a ``sale'' of "personal information." If you've visited our digital properties within the past 12 months and you've seen ads, under California law personal information about you may have been ``sold'' to our advertising partners for their own use. California residents have the right to opt out of the ``sale'' of personal information, and we've made it easy for anyone to stop the information transfers that might be considered such a ``sale'' from our website or mobile app.

\medskip
\textbf{How to Opt Out of the Sale of Your Information}

For our website, click on the ``Do Not Sell My Personal Information'' button. For our mobile app, look under Account, then Extras, then Do Not Sell My Personal Information and make the selection.

\medskip
\textbf{What will happen?}

After you click the ``Do Not Sell My Personal Information'' button:
\begin{itemize}
    \item \textbf{Opt-out cookie}: An opt-out cookie will be placed and stored on your browser, preventing personal information from being made available from this website to advertising partners for their own use, independent of Company XYZ. 
\end{itemize}

\textbf{A few things to keep in mind.}
\begin{itemize}
    \item \textbf{Ads and interest-based advertising}: You will still see ads. You are not opted out of interest-based advertising. To opt out of interest-based advertising, visit www.aboutads.info/choices.
    
    \item \textbf{If you delete or clear cookies, you'll need to click the button again the next time you visit.} If you delete or clear your cookies, that will delete our opt-out cookie and you will need to opt out again.
\end{itemize}

\hrulefill
\begin{itemize}
    \item As a California resident, if you opt out of the sale of your personal information as described, is Company XYZ allowed to provide \textbf{your full name and mailing address} to \textbf{an advertising company} so that the advertising company can display advertisements to you on other websites, \textbf{without receiving revenue in return} from this advertising company?
    \begin{itemize}
        \item Definitely allowed / Probably allowed / I'm really not sure / Probably not allowed / Definitely not allowed
    \end{itemize}
    
    \item As a California resident, if you opt out of the sale of your personal information as described, is Company XYZ allowed to provide your \textbf{full name and mailing address} to \textbf{a separately-owned electronics retailer} so that the two companies can provide joint product offers to you, \textbf{without receiving revenue in return} from this partner company?
    \begin{itemize}
        \item Definitely allowed / Probably allowed / I'm really not sure / Probably not allowed / Definitely not allowed
    \end{itemize}
    
    \item Which of the following categories of personal information did the previous question involve? [\textit{Attention check}]
\end{itemize}
\hrulefill
\medskip

\textbf{Please read the following section of text from the privacy policy of Company XYZ, an electronics retailer, and answer the questions below.}

\textit{Note that hyperlinks have been removed from this excerpt.}

\medskip
You have certain choices regarding our use and disclosure of your personal information, as described below.
\begin{itemize}
    \item \textbf{Deletion}: You have the right to request that we delete certain personal information we have collected from you. Exceptions apply.
\end{itemize}

\textbf{How to Submit a Request}
\begin{itemize}
    \item Submit a deletion request, or call us at 1-555-STAN-XYZ.
\end{itemize}

\textbf{Verifying Requests.}

To help protect your privacy and maintain security, we will take steps to verify your identity before granting you access to your personal information or complying with your request. If you have an account with us, we may verify your identity by requiring you to sign in to your account. If you request access to or deletion of your personal information and do not sign in to an account with us, we require you to provide the following information: name, email address, phone number, and postal address. In addition, if you do not have an account and you ask us to provide you with specific pieces of personal information, we reserve the option to require you to sign a declaration under penalty of perjury that you are the consumer whose personal information is the subject of the request.

\hrulefill
\begin{itemize}
    \item As a California resident, if you request the deletion of your personal information as described, is Company XYZ required to direct \textbf{any marketing company that has received your information, either by sharing or selling, from Company XYZ} to delete information about your activity on Company XYZ's websites?
    \begin{itemize}
        \item Definitely required / Probably required / I'm really not sure / Probably not required / Definitely not required
    \end{itemize}
    
    \item As a California resident, if you request the deletion of your personal information as described, is Company XYZ required to direct \textbf{any marketing company that provides advertising consultation services to Company XYZ} (that is, acting in the capacity of a service provider) to delete information about your activity on Company XYZ's websites?
    \begin{itemize}
        \item Definitely required / Probably required / I'm really not sure / Probably not required / Definitely not required
    \end{itemize}
\end{itemize}
\hrulefill

\subsubsection{Privacy Policy 7: Google} 
\hspace*{\fill} \\
\textbf{Please read the following section of text from the privacy policy of Company XYZ, a technology company that provides web search, web mail, cloud storage, video hosting, and other products and services, and answer the questions below.}

\textit{Note that hyperlinks have been removed from this excerpt.}

\medskip
The CCPA requires a description of data practices using specific categories. This table uses these categories to organize the information in this Privacy Policy. [\textit{Table reproduced to the side}]

\hrulefill
\begin{itemize}
    \item As a California resident, if you make purchases through a website owned by Company XYZ (for example, if you purchase apps on the XYZ App Store and pay with PayPal), is Company XYZ allowed to share your \textbf{past purchase history} with a payment service provider (such as PayPal)?
    \begin{itemize}
        \item Definitely allowed / Probably allowed / I'm really not sure / Probably not allowed / Definitely not allowed
    \end{itemize}
    
    \item As a California resident, if you make purchases through a website owned by Company XYZ (for example, if you purchase apps on the XYZ App Store and pay with PayPal), is Company XYZ allowed to share your \textbf{inferred ethnicity} (inferred based on your purchase history) with a payment service provider (such as PayPal)?
    \begin{itemize}
        \item Definitely allowed / Probably allowed / I'm really not sure / Probably not allowed / Definitely not allowed
    \end{itemize}
    
    \item What was the category of the website in the privacy statement you just read? [\textit{Attention check}]
\end{itemize}
\hrulefill

\xentrystretch{-1}
\begin{xtabular}{|>{\raggedright\arraybackslash}p{\dimexpr.5\columnwidth-3\tabcolsep}|>{\raggedright\arraybackslash}p{\dimexpr.5\columnwidth-3\tabcolsep}|}
    \hline
    \textbf{Categories of personal information we collect} & \textbf{Parties with whom information may be shared} \\
    \hline
    \begin{itemize}[leftmargin=1em,topsep=0pt]
        \item \textbf{Identifiers} such as your name, phone number, and address, as well as unique identifiers tied to the browser, application, or device you're using.
        
        \item \textbf{Demographic information}, such as your age, gender and language.
        
        \item \textbf{Commercial information} such as your payment information and a history of purchases you make on Company XYZ's services.
        
        \item \textbf{Internet, network, and other activity information} such as your search terms; views and interactions with content and ads; XYZ Browser browsing history you've synced with your XYZ Account; information about the interaction of your apps, browsers, and devices with our services (like IP address, crash reports, and system activity); and activity on third-party sites and apps that use our services. You can review and control activity data stored in your XYZ Account in My Activity.

        \item \textbf{Geolocation data}, such as may be determined by GPS, IP address, and other data from sensors on or around your device, depending in part on your device and account settings. Learn more about Company XYZ's use of location information.

        \item \textbf{Other information you create or provide}, such as the content you create, upload, or receive (like photos and videos or emails, docs and spreadsheets). XYZ Dashboard allows you to manage information associated with specific products.

        \item \textbf{Inferences} drawn from the above, like your ads interest categories.
    \end{itemize} &
    \begin{itemize}[leftmargin=1em,topsep=0pt]
        \item \textbf{Other people with whom you choose to share your information}, like docs or photos, and videos or comments on XYZ Videos.

        \item \textbf{Third parties to whom you consent to sharing your information}, such as services that integrate with Company XYZ's services. You can review and manage third party apps and sites with access to data in your XYZ Account.

        \item \textbf{Service providers}, trusted businesses or persons that process information on Company XYZ's behalf, based on our instructions and in compliance with our Privacy Policy and any other appropriate confidentiality and security measures.

        \item \textbf{Law enforcement or other third parties}, for the legal reasons described in Sharing your information.
    \end{itemize} \\
    \hline
\end{xtabular}

\newpage
\medskip
\textbf{Please read the following section of text from the privacy policy of Company XYZ, a technology company that provides web search, web mail, cloud storage, video hosting, and other products and services, and answer the questions below.}

\textit{Note that hyperlinks have been removed from this excerpt.}

This Privacy Policy is designed to help you understand how Company XYZ handles your information:
\begin{itemize}
    \item We explain the categories of information Company XYZ collects and the sources of that information in Information Company XYZ collects.
    
    \item We explain how Company XYZ uses information in Why Company XYZ collects data.
    
    \item We explain when Company XYZ may share information in Sharing your information. Company XYZ does not sell your personal information.
\end{itemize}

We describe the choices you have to manage your privacy and data across Company XYZ's services in Your privacy controls. You can exercise your rights by using these controls, which allow you to access, review, update and delete your information, as well as export and download a copy of it. When you use them, we'll validate your request by verifying that you're signed in to your XYZ Account. If you have questions or requests related to your rights under the California Consumer Privacy Act (CCPA), you (or your authorized agent) can also contact Company XYZ.

\hrulefill
\begin{itemize}
    \item Given that you are a California resident, could Company XYZ be allowed to provide your \textbf{full name and mailing address} to an advertising company, \textbf{without receiving revenue in return} from this advertising company?
    \begin{itemize}
        \item Definitely allowed / Probably allowed / I'm really not sure / Probably not allowed / Definitely not allowed
    \end{itemize}
    
    \item Given that you are a California resident, if Company XYZ provides your \textbf{full name and mailing address} to an advertising company, \textbf{without receiving revenue in return} from this advertising company, would you \textbf{have an option to opt-out} of this sharing?
    \begin{itemize}
        \item Definitely yes / Probably yes / I'm really not sure / Probably no / Definitely no
    \end{itemize}
    
    \item Which of the following categories of personal information did the previous question involve? [\textit{Attention check}]
\end{itemize}
\hrulefill
\medskip

\textbf{Please read the following section of text from the privacy policy of Company XYZ, a technology company that provides web search, web mail, cloud storage, video hosting, and other products and services, and answer the questions below.}

\textit{Note that hyperlinks have been removed from this excerpt.}

\medskip
To delete your information, you can:
\begin{itemize}
    \item Delete your content from specific Company XYZ services

    \item Search for and then delete specific items from your account using My Activity

    \item Delete specific Company XYZ products, including your information associated with those products

    \item Delete your entire XYZ Account
\end{itemize}

We retain the data we collect for different periods of time depending on what it is, how we use it, and how you configure your settings:
\begin{itemize}
    \item Some data you can delete whenever you like, such as the content you create or upload. You can also delete activity information saved in your account, or choose to have it deleted automatically after a set period of time.

    \item Other data is deleted or anonymized automatically after a set period of time, such as advertising data in server logs.

    \item We keep some data until you delete your XYZ Account, such as information about how often you use our services.

    \item And some data we retain for longer periods of time when necessary for legitimate business or legal purposes, such as security, fraud and abuse prevention, or financial record-keeping.
\end{itemize}

When you delete data, we follow a deletion process to make sure that your data is safely and completely removed from our servers or retained only in anonymized form. We try to ensure that our services protect information from accidental or malicious deletion. Because of this, there may be delays between when you delete something and when copies are deleted from our active and backup systems.

\medskip
You can read more about Company XYZ's data retention periods, including how long it takes us to delete your information.

\hrulefill
\begin{itemize}
    \item As a California resident, if you request the deletion of your Company XYZ account, is Company XYZ required to direct \textbf{any marketing company that has received your information, either by sharing or selling, from Company XYZ} to delete information about your activity on Company XYZ's websites?
    \begin{itemize}
        \item Definitely required / Probably required / I'm really not sure / Probably not required / Definitely not required
    \end{itemize}
    
    \item As a California resident, if you request the deletion of your Company XYZ account, is Company XYZ required to direct \textbf{any marketing company that provides advertising consultation services to Company XYZ} (that is, acting in the capacity of a service provider) to delete information about your activity on Company XYZ's websites?
    \begin{itemize}
        \item Definitely required / Probably required / I'm really not sure / Probably not required / Definitely not required
    \end{itemize}
\end{itemize}
\hrulefill

\subsubsection{Privacy Policy 8: Microsoft}
\hspace*{\fill} \\
\textbf{Please read the following section of text from the privacy policy of Company XYZ, a technology company that provides software, web search, web mail, cloud storage, and other products and services, and answer the questions below.}

\medskip
\textbf{Personal Information Processing.} In the bulleted list below, we outline the categories of personal data we collect, the sources of the personal data, our purposes of processing, and the categories of third-party recipients with whom we share the personal data. For a description of the data included in each category, please see the Personal data we collect section.

\medskip
\textbf{Categories of Personal Data}
\begin{itemize}
    \item \textbf{Demographic data}
    \begin{itemize}
        \item \textbf{Sources of personal data}: Interactions with users and purchases from data brokers
        \item \textbf{Purposes of Processing} (Collection and Sharing with Third Parties): Provide and personalize our products; product development; help, secure, and troubleshoot; and marketing
        \item \textbf{Recipients}: Service providers and user-directed entities
    \end{itemize}
    
    \item \textbf{Payment data}
    \begin{itemize}
        \item \textbf{Sources of personal data}: Interactions with users and financial institutions
        \item \textbf{Purposes of Processing} (Collection and Sharing with Third Parties): Transact commerce; process transactions; fulfill orders; help, secure, and troubleshoot; and detect and prevent fraud
        \item \textbf{Recipients}: Service providers and user-directed entities
    \end{itemize}
\end{itemize}

\hrulefill
\begin{itemize}
    \item As a California resident, if you make purchases through a website owned by Company XYZ (for example, if you purchase a game through Company XYZ's Games Store and pay with PayPal), is Company XYZ allowed to share your \textbf{past purchase history} with a payment service provider (such as PayPal)?
    \begin{itemize}
        \item Definitely allowed / Probably allowed / I'm really not sure / Probably not allowed / Definitely not allowed
    \end{itemize}
    
    \item As a California resident, if you make purchases through a website owned by Company XYZ (for example, if you purchase a game through Company XYZ's Games Store and pay with PayPal), is Company XYZ allowed to share your \textbf{inferred ethnicity} (inferred based on your purchase history) with a payment service provider (such as PayPal)?
    \begin{itemize}
        \item Definitely allowed / Probably allowed / I'm really not sure / Probably not allowed / Definitely not allowed
    \end{itemize}
    
    \item What was the category of the website in the privacy statement you just read? [\textit{Attention check}]
\end{itemize}
\hrulefill
\medskip

\textbf{Please read the following section of text from the privacy policy of Company XYZ, a technology company that provides software, web search, web mail, cloud storage, and other products and services, and answer the questions below.}

\textit{Note that hyperlinks have been removed from this excerpt.}

\medskip
\textbf{Sale.} We do not sell your personal data. So, we do not offer an opt-out to the sale of personal data.

\medskip
\textbf{Rights.} You have the right to request that we (i) disclose what personal data we collect, use, disclose, and sell and (ii) delete your personal data. You may make these requests yourself or through an authorized agent. If you use an authorized agent, we provide your agent with detailed guidance on how to exercise your California Consumer Privacy Act (CCPA) rights.

\medskip
If you have a Company XYZ account, you must exercise your rights through the Company XYZ privacy dashboard, which requires you to log in to your Company XYZ account. If you have an additional request or questions after using the dashboard, you may contact Company XYZ at the address in the How to contact us section, use our web form, or call our US toll free number 1.555.273.7826. If you do not have an account, you may exercise your rights by contacting us as described above. We may ask for additional information, such as your country of residence, email address, and phone number, to validate your request before honoring the request.

\medskip
You have a right not to receive discriminatory treatment if you exercise your CCPA rights. We will not discriminate against you if you exercise your CCPA rights.

\hrulefill
\begin{itemize}
    \item Given that you are a California resident, could Company XYZ be allowed to provide your \textbf{full name and mailing address} to an advertising company, \textbf{without receiving revenue in return} from this advertising company?
    \begin{itemize}
        \item Definitely allowed / Probably allowed / I'm really not sure / Probably not allowed / Definitely not allowed
    \end{itemize}
    
    \item Given that you are a California resident, if Company XYZ provides your \textbf{full name and mailing address} to an advertising company, \textbf{without receiving revenue in return} from this advertising company, would you \textbf{have an option to opt-out} of this sharing?
    \begin{itemize}
        \item Definitely yes / Probably yes / I'm really not sure / Probably no / Definitely no
    \end{itemize}
    
    \item Which of the following categories of personal information did the previous question involve? [\textit{Attention check}]
\end{itemize}
\hrulefill
\medskip

\textbf{Please read the following section of text from the privacy policy of Company XYZ, a technology company that provides software, web search, web mail, cloud storage, and other products and services, and answer the questions below.}

\textit{Note that hyperlinks have been removed from this excerpt.}

\medskip
\textbf{Rights.} You have the right to request that we (i) disclose what personal data we collect, use, disclose, and sell and (ii) delete your personal data. You may make these requests yourself or through an authorized agent. If you use an authorized agent, we provide your agent with detailed guidance on how to exercise your California Consumer Privacy Act (CCPA) rights.

\medskip
If you have a Company XYZ account, you must exercise your rights through the Company XYZ privacy dashboard, which requires you to log in to your Company XYZ account. If you have an additional request or questions after using the dashboard, you may contact Company XYZ at the address in the How to contact us section, use our web form, or call our US toll free number 1.555.273.7826. If you do not have an account, you may exercise your rights by contacting us as described above. We may ask for additional information, such as your country of residence, email address, and phone number, to validate your request before honoring the request.

\medskip
You have a right not to receive discriminatory treatment if you exercise your CCPA rights. We will not discriminate against you if you exercise your CCPA rights.

\hrulefill
\begin{itemize}
    \item As a California resident, if you request the deletion of your Company XYZ account, is Company XYZ required to direct \textbf{any marketing company that has received your information, either by sharing or selling, from Company XYZ} to delete information about your activity on Company XYZ's websites?
    \begin{itemize}
        \item Definitely required / Probably required / I'm really not sure / Probably not required / Definitely not required
    \end{itemize}
    
    \item As a California resident, if you request the deletion of your Company XYZ account, is Company XYZ required to direct \textbf{any marketing company that provides advertising consultation services to Company XYZ} (that is, acting in the capacity of a service provider) to delete information about your activity on Company XYZ's websites?
    \begin{itemize}
        \item Definitely required / Probably required / I'm really not sure / Probably not required / Definitely not required
    \end{itemize}
\end{itemize}
\hrulefill

\subsubsection{Privacy Policy 9: Netflix} 
\hspace*{\fill} \\
\textbf{Please read the following section of text from the privacy policy of Company XYZ, a streaming service, and answer the questions below.}

\medskip
\uline{Categories of CCPA personal information disclosed for a business purpose}

We disclose the categories of California Consumer Privacy Act (CCPA) personal information listed below for business purposes. (Please see the Disclosure of Information section of our Privacy Statement for additional details that may be of interest to you.)

\begin{itemize}
    \item \textbf{Identifiers}: We may disclose identifiers for business purposes with the following categories of third parties: Service Providers, Partners, an entity engaged in a business transfer, law enforcement, courts, governments and regulatory agencies.

    \item \textbf{Characteristics of protected classifications under California or federal law}: We may disclose these types of characteristics for business purposes with the following categories of third parties: Service Providers, an entity engaged in a business transfer/merger, law enforcement, courts, governments and regulatory agencies.

    \item \textbf{Commercial information}: We may disclose commercial information for business purposes with the following categories of third parties: Service Providers, Partners, an entity engaged in a business transfer/merger, law enforcement, courts, governments and regulatory agencies.

    \item \textbf{Internet or other electronic network activity information}: We may disclose these types of information for business purposes with the following categories of third parties: Service Providers, Partners, an entity engaged in a business transfer/merger, law enforcement, courts, governments and regulatory agencies.

    \item \textbf{Geolocation data}: We may disclose geolocation data for business purposes with the following categories of third parties: Service Providers, Partners, an entity engaged in a business transfer/merger, law enforcement, courts, governments and regulatory agencies.

    \item \textbf{Inferences}: We may disclose these types of data for business purposes with the following categories of third parties: an entity engaged in a business transfer/merger.
\end{itemize}

\hrulefill
\begin{itemize}
    \item As a California resident, if you make purchases through a website owned by Company XYZ (for example, if you make a monthly payment for your XYZ Streaming Service subscription with PayPal), is Company XYZ allowed to share your \textbf{past purchase history} with a payment service provider (such as PayPal)?
    \begin{itemize}
        \item Definitely allowed / Probably allowed / I'm really not sure / Probably not allowed / Definitely not allowed
    \end{itemize}
    
    \item As a California resident, if you make purchases through a website owned by Company XYZ (for example, if you make a monthly payment for your XYZ Streaming Service subscription with PayPal), is Company XYZ allowed to share your \textbf{inferred ethnicity} (inferred based on your purchase history) with a payment service provider (such as PayPal)?
    \begin{itemize}
        \item Definitely allowed / Probably allowed / I'm really not sure / Probably not allowed / Definitely not allowed
    \end{itemize}
    
    \item What was the category of the website in the privacy statement you just read? [\textit{Attention check}]
\end{itemize}
\hrulefill
\medskip

\textbf{Please read the following section of text from the privacy policy of Company XYZ, a streaming service, and answer the questions below.}

\medskip
\uline{Your rights under the California Consumer Privacy Act (CCPA)}

\begin{itemize}
    \item You have the right to request that we disclose: what categories and specific pieces of CCPA personal information have been collected about you; the categories of sources from which CCPA personal information are collected; our business or commercial purpose for collecting, using, or disclosing CCPA personal information; the categories of third parties with whom we share CCPA personal information; the categories of CCPA personal information we have disclosed about you for a business purpose. We do not sell personal information.

    \item You have a right to receive a copy of the specific CCPA personal information we have collected about you.

    \item You have a right to deletion of your CCPA personal information, subject to exceptions under the CCPA.

    \item You have a right not to receive discriminatory treatment for exercising any of your CCPA rights. We will not discriminate against you based on your exercise of any of your CCPA rights.
\end{itemize}

You can assert these rights only where we receive a verified request from you.

\hrulefill
\begin{itemize}
    \item Given that you are a California resident, could Company XYZ be allowed to provide your \textbf{full name and mailing address} to an advertising company, \textbf{without receiving revenue in return} from this advertising company?
    \begin{itemize}
        \item Definitely allowed / Probably allowed / I'm really not sure / Probably not allowed / Definitely not allowed
    \end{itemize}
    
    \item Given that you are a California resident, if Company XYZ provides your \textbf{full name and mailing address} to an advertising company, \textbf{without receiving revenue in return} from this advertising company, would you \textbf{have an option to opt-out} of this sharing?
    \begin{itemize}
        \item Definitely yes / Probably yes / I'm really not sure / Probably no / Definitely no
    \end{itemize}
    
    \item Which of the following categories of personal information did the previous question involve? [\textit{Attention check}]
\end{itemize}
\hrulefill
\medskip

\textbf{Please read the following section of text from the privacy policy of Company XYZ, a streaming service, and answer the questions below.}

\textit{Note that hyperlinks have been removed from this excerpt.}

\medskip
\textbf{Your Information and Rights}

You can request access to your personal information, or correct or update out-of-date or inaccurate personal information we hold about you. You may also request that we delete personal information that we hold about you.

\medskip
If you are the account owner, to download a copy of your personal information go to: \url{www.xyzstreaming.com/account/getmyinfo} (you must be signed in to access the ``Account'' section), and follow the instructions.

\medskip
For other requests, or if you have any other question regarding our privacy practices, please contact our Data Protection Officer/Privacy Office at privacy@xyzstreaming.com. For information about deletion, removal and retention of information, please reference this help article: \newline help.xyzstreaming.com/node/100625. We respond to all requests we receive from individuals wishing to exercise their data protection rights in accordance with applicable data protection laws.

\medskip
We may reject requests that are unreasonable or not required by law, including those that would be extremely impractical, could require disproportionate technical effort, or could expose us to operational risks such as free trial fraud. We may retain information as required or permitted by applicable laws and regulations, including to honor your choices, for our billing or records purposes and to fulfill the purposes described in this Privacy Statement. We take reasonable measures to destroy or de-identify personal information in a secure manner when it is no longer required.

\hrulefill
\begin{itemize}
    \item As a California resident, if you request the deletion of your Company XYZ account, is Company XYZ required to direct \textbf{any marketing company that has received your information, either by sharing or selling, from Company XYZ} to delete information about your activity on Company XYZ's websites?
    \begin{itemize}
        \item Definitely required / Probably required / I'm really not sure / Probably not required / Definitely not required
    \end{itemize}
    
    \item As a California resident, if you request the deletion of your Company XYZ account, is Company XYZ required to direct \textbf{any marketing company that provides advertising consultation services to Company XYZ} (that is, acting in the capacity of a service provider) to delete information about your activity on Company XYZ's websites?
    \begin{itemize}
        \item Definitely required / Probably required / I'm really not sure / Probably not required / Definitely not required
    \end{itemize}
\end{itemize}
\hrulefill

\subsection{Post-survey questions}
\subsubsection{Demographic questions}
\begin{itemize}
    \item How did you find the surveys related to privacy or security that you completed in the past year? [\textit{Attention check; respondents indicate whether they had completed any such surveys in the past year in the screener}]
    
    \item Please select the category which best describes your age range.
    
    \item What is your gender identity?
    
    \item Please select the category which best describes your marital status.
    
    \item What is the highest level of school you have completed or the highest degree you have received?
    
    \item Which of the following best describes your employment status?
    
    \item Please select the category which best describes your field of study or employment.
\end{itemize}

\subsubsection{Privacy and security questions}
Each statement in the following section describes how a person might feel about the privacy policy you just read, as well as the use of security measures. Examples of security measures include laptop or tablet passwords, spam email reporting tools, software updates, secure web browsers, fingerprint IDs, and anti-virus software. Please indicate the degree to which you agree or disagree with each statement in the following section. For each statement, make your choice in terms of how you feel \textbf{right now}, not what you have felt in the past or how you would like to feel. There are no wrong answers.

\begin{itemize}
    \item The privacy policy excerpts I read contained provisions for unexpected uses of my personal information.
    
    \item I would be more likely to read or skim the privacy policies on websites that I visit in the future.
    
    \item I believe that the potential risks in the privacy policy excerpts I read would be adequately mitigated by privacy regulations in my jurisdiction, such as the California Consumer Privacy Act (CCPA).
\end{itemize}

[\textit{The ensuing questions are derived from the SA-6 self-report measure of security attitudes \cite{Faklaris2019}.}]

\begin{itemize}
    \item I always pay attention to expert advice about the steps that I need to take to keep my personal information, including online data and accounts, safe.
    
    \item Generally speaking, I follow a routine of security practices diligently.
    
    \newpage
    \item I seek out opportunities to learn about security measures that are relevant to me.
    
    \item I am extremely motivated to take all the steps needed to keep my personal information, including online data and accounts, safe.
    
    \item I often am interested in articles about security threats.
    
    \item I am extremely knowledgeable about all the steps needed to keep my personal information, including online data and accounts, safe.
\end{itemize}
}

\section{Power analysis details}
\label{sec:appendix:power}
In a pilot of 15 respondents, we saw that 40\% of questions were answered correctly and 60\% of questions were answered incorrectly. Under the null hypothesis, we treated these as the true answer correctness proportions for each question-policy combination. For \textbf{Q1}, \textbf{Q2}, \textbf{Q5}, and \textbf{Q6}, under the alternate hypothesis, we expected a significant deviation in these proportions --- the equivalent of deviating to 55\% correct and 45\% incorrect for at least six out of nine policies, corresponding to the less ambiguous policies in our survey (Section \ref{sec:survey-bg:policy}). This yielded an effect size of 0.25. With $(2 - 1)(9 - 1) = 8$ degrees of freedom, we needed 364 respondents to achieve a significance level of $\alpha = 0.05$ and a power of $1 - \beta = 0.95$. 

For \textbf{Q3} and \textbf{Q4}, for which respondents were divided approximately evenly between policies, there are $(2 - 1)(3 - 1) = 2$ degrees of freedom. A third of the overall sample size ($\approx 121$) can detect an effect size of 0.357 for the same $\alpha$ and $1 - \beta$ (corresponding to slightly larger deviations to 61.5\% and 38.5\% for two out of three policies). While our sample size does not allow us to detect smaller effects, we believe that it provides a reasonable trade-off between statistical significance and practical feasibility. 

\begin{table*}[hb]
\section{Summary of survey results}
\label{sec:appendix:correct-chisq}
\medskip
    \centering
    \begin{tabularx}{0.75\textwidth}{@{}XXXXX@{}}
        \toprule
        Question & $n$ & \% correct & $\chi^2$ statistic & $p$-value \\
        \midrule
        \textbf{Q1} & 364 & 80.77\% & \textbf{19.762} & \textbf{0.0113} \\
        \textbf{Q2} & 364 & 63.74\% & \textbf{16.816} & \textbf{0.0321} \\
        \midrule
        \textbf{Q3} All & 364 & 44.78\% & \textbf{162} & \textbf{1.944} $\mathbf{\times 10^{-6}}$ \\
        \textbf{Q3} Sharing & 122 & 38.52\% & \textbf{3.881} & \textbf{0.0488} \\
        \textbf{Q3} Advertising & 121 & 30.58\% & 3.515 & 0.0608 \\
        \textbf{Q3} No sale & 121 & 65.29\% & 2.196 & 0.3335 \\
        \midrule
        \textbf{Q4} Sharing & 122 & 80.33\% & 0.767 & 0.3811 \\
        \textbf{Q4} Advertising & 121 & 54.55\% & 0.594 & 0.4410 \\
        \textbf{Q4} No sale & 121 & 61.16\% & \textbf{14.233} & \textbf{0.0008} \\
        \midrule
        \textbf{Q5} & 364 & 57.69\% & \textbf{26.130} & \textbf{0.0010} \\
        \textbf{Q6} & 364 & 35.44\% & \textbf{26.003} & \textbf{0.0010} \\
        \bottomrule
    \end{tabularx}
    \captionsetup{type=table}
    \captionof{table}{Results of the $\chi^2$ independence tests that we performed for each question. Statistics that are significant at the 0.05 level are bolded. $n$ is the number of survey respondents who provided answers for the given question type.}
    \label{tab:chisq}
\end{table*}
\phantom{ }

\end{document}